\documentclass[journal,10pt]{IEEEtran}
\usepackage{subfigure}
\PassOptionsToPackage{subfigure}{tocloft}
\usepackage{CJK}
\usepackage{algorithm}
\usepackage{algpseudocode}
\usepackage{psfrag}
\usepackage{graphicx}
\usepackage{epstopdf}
\usepackage{multirow}
\usepackage{cite}
\usepackage{arydshln}
\usepackage{enumerate}
\usepackage{bbm}
\usepackage{bm}
\usepackage{extarrows}

\usepackage{float}
\usepackage{stfloats}

\usepackage{multirow}
\usepackage[dvipsnames]{xcolor}

\usepackage{amsthm}
\usepackage{amsmath}
\usepackage{amssymb}

\theoremstyle{plain}

\usepackage{xcolor}

\usepackage{changes}

\usepackage{etoolbox}

\usepackage{physics}

\let\mybibitem\bibitem
\renewcommand{\bibitem}[1]{%
\ifstrequal{#1}{Alex2023}{\mybibitem{#1}}
{\color{black}\mybibitem{#1}
}%
}

\usepackage{lipsum}

\title{Channel Estimation and Analog Precoding for\\ Pixel-based Fluid-Antenna-Assisted Multiuser \\ MIMO-OFDM Systems}
\author{Huayan Guo, \IEEEmembership{Member, IEEE}, Jichen Zhang, \IEEEmembership{Graduate Student Member, IEEE}, Junhui Rao, \IEEEmembership{Member, IEEE},\\
  Ross Murch, \IEEEmembership{Fellow, IEEE}, and Vincent~K.~N.~Lau, \IEEEmembership{Fellow, IEEE}
\thanks{This work was supported in part by the National Natural Science Foundation of China under Grant 62101472, in part by the Research Grants Council of Hong Kong under Project 16214122, and in part by the Research Grants Council
under the Areas of Excellence scheme grant  AoE/E-101/23-N (Corresponding author: Vincent K. N. Lau.)}
\thanks{Huayan Guo, Vincent K. N. Lau, Junhui Rao, and Ross Murch  are with the Department of Electronic and Computer Engineering, The Hong Kong University of Science and Technology, Clear Water Bay, Kowloon, Hong Kong 999077 (e-mail: { eeguohuayan@ust.hk; jzhangiq@connect.ust.hk; jraoaa@connect.ust.hk; eermurch@ust.hk; eeknlau@ece.ust.hk}).}
}

\begin{document}
\maketitle
\IEEEpeerreviewmaketitle
\begin{abstract}
Pixel-based fluid antennas provide enhanced multiplexing gains and quicker radiation pattern switching than traditional designs. However, this innovation introduces challenges for channel estimation and analog precoding due to the state-non-separable channel response problem. This paper explores a multiuser MIMO-OFDM system utilizing pixel-based fluid antennas, informed by measurements from a real-world prototype. We present a sparse channel recovery framework for uplink channel sounding, employing an approximate separable channel response model with DNN-based antenna radiation functions. We then propose two low-complexity channel estimation algorithms that leverage orthogonal matching pursuit and variational Bayesian inference to accurately recover channel responses across various scattering cluster angles. These estimations enable the prediction of composite channels for all fluid antenna states, leading to an analog precoding scheme that optimally selects switching states for different antennas. Our simulation results indicate that the proposed approach significantly outperforms several baseline methods, especially in high signal-to-noise ratio environments with numerous users.
\end{abstract}

\begin{IEEEkeywords}
Fluid antenna, movable antenna, next-generation reconfigurable antenna, state-non-separable channel response, channel estimation
\end{IEEEkeywords}

\section{Introduction}\label{sec:introduction}
Massive multiple-input multiple-output (MIMO) technology is a key enabler for the next generation of wireless networks. To achieve more flexible beam and higher multiplexing gain, the concept of a fluid antenna system (FAS) has recently been proposed for massive MIMO \cite{FAS2020TWC@KKWong,FAS2023WCL@KKWong,Fluid2025Survey}. This system can arbitrarily adjust the intrinsic properties of antennas, including shape, position, radiation pattern, operating frequency, and other characteristics. However, examples of FAS designs are currently limited, with most existing research primarily focusing on mechanically movable antenna systems \cite{movable2025Survey}, where antenna positions are optimized to enhance MIMO performance \cite{MovableOPT2023TWC,MovableOPT2025JSAC,sumrateMFAS2024CL,MUMFASopt2024WCL}. Unfortunately, adjustments based on physical movement occur relatively slowly compared to packet transmission rates (about 1 ms per packet), significantly constraining overall performance. To address this limitation, a pixel-based fluid antenna is proposed in \cite{Pixelantenna2025}, achieving spatial sampling equivalent to port movement with $\mu s$-level switching speeds by adjusting the  radiation pattern of each antenna across 12 candidate states. The remaining task involves configuring the states of the fluid antenna array (FAA), a process also referred to as analog precoding.

\begin{figure}
[!t]
\centering
\includegraphics[width=.90\columnwidth]{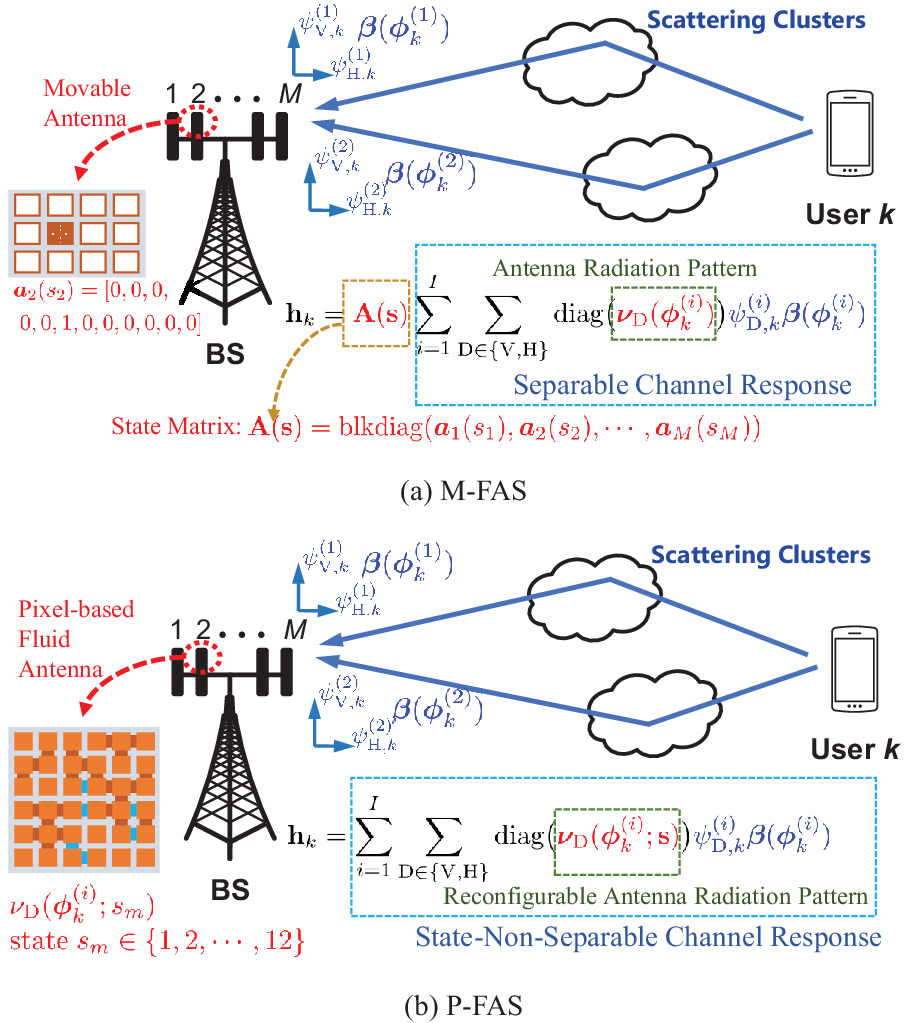}
\caption{Illustration of the fluid-antenna-assisted downlink MISO systems: (a) M-FAS; (b) P-FAS.}
\label{FAS_illustration}
\vspace{-1.5em}
\end{figure}

Channel estimation is essential for precoder design to fully achieve the performance gains provided by massive antennas. Most existing research focuses on channel estimation in the movable-antenna-based FAS (M-FAS). In particular, the movable antenna can be viewed as a generalized hybrid beamforming (HBF) structure, where all possible antenna positions create a virtual antenna sub-array connected to a radio-frequency (RF) chain, making the antenna placement an antenna selection problem within the sub-array \cite{movableCE2024WCL,movableCE2023CL,movableCEBayesian2024TWC,movableCEoversample2025TWC}. Therefore, all sparse channel recovery algorithms developed for HBF can be applied to the M-FAS \cite{sparseCE2010magazine,burseLASSO2016TWC,MarkovpriorTVT2017}. Specifically, the effective channel from the digital RF chains of the transmitter to the receiver antennas can be represented as the product of the analog precoder and the wireless propagation channel, as shown in {\figurename~\ref{FAS_illustration}(a)}. This {\em separable channel response} property enables the prediction of the effective channel for any analog precoder based on an estimated propagation channel. Consequently, estimating the propagation channel is sufficient for optimizing the analog precoder.

Unfortunately, the pixel-based FAS (P-FAS) encounters a {\em state-non-separable channel  response} issue, which complicates channel estimation. Specifically, although the effective channel can be expressed as the product of the array response  and the wireless propagation channel, the array response matrix is a non-separable function of the FAA states and the angles of departure of the radio scattering paths \cite{Pixelantenna2025}, as illustrated in {\figurename~\ref{FAS_illustration}(b)}. Consequently, channel estimation for a specific FAA state configuration cannot be directly utilized to predict the effective channel at other configurations, resulting in the failure of the traditional analog precoder design flow. In \cite{kernelCE2024TC,FASCERay2023CL}, this issue is addressed by estimating the angles, delays, and ray responses of all the scattering paths, allowing for the prediction of the array response for any state configuration. However, this algorithm is only applicable when the number of rays is very small (such as 7 in \cite{kernelCE2024TC} and 5 in \cite{FASCERay2023CL}), which limits its utility for sub-7 GHz scenarios. In \cite{greedyoptStatisticCE2023WSA,optStatistic2023CL}, instead of estimating the instantaneous channel, the angles and average powers of the dominant scattering paths are estimated from sufficient historical channel observations, enabling the construction of statistical knowledge about the effective channel for any state configuration. However, this algorithm may not perform well when the scattering clusters of different users are closely located, as the statistical channel information may not effectively suppress interference.

In this paper, we investigate a time-division duplexing (TDD) multiuser MIMO-OFDM system utilizing pixel-based fluid antennas (proposed in \cite{Pixelantenna2025}) at the BSs. To tackle the issue of the {\em state-non-separable channel response}, we introduce an {\em approximate separable channel response} model, leveraging the slowly varying radiation pattern function of the pixel-based fluid antenna concerning scattering ray angles. We formulate the uplink channel estimation problem as a sparse channel recovery problem over a measurement matrix associated with a high-resolution angular grid. However, estimating the channel within this model is challenging due to the extremely singular measurement matrix resulting from the high-resolution antenna-angular transformation, which degrades the performance of approximate-message-passing (AMP)-based algorithms \cite{AMP,VAMP,OAMP,TurboOMP2015SPL}. Furthermore, sparse channel recovery over the high-resolution angular basis incurs considerable computational complexity. Analog precoding also poses challenges due to the discrete nature of antenna state configurations. While some heuristic methods have been proposed \cite{greedyoptStatisticCE2023WSA,greedyopt2019TWC}, their performance remains suboptimal. This paper presents low-complexity algorithms for both uplink channel estimation and downlink analog precoding that demonstrate improved performance. The following summarizes the key contributions of this work:

\begin{itemize}
\item  {\bf {A Holistic Solution for the P-FAS based on Measurements from a Real-World Prototype}}:
We provide a holistic solution for a TDD P-FAS, encompassing uplink channel estimation and downlink analog  precoding. Both the algorithm design and performance evaluation are based on measurements from a real-world P-FAS prototype reported in \cite{Pixelantenna2025}. Moreover, to facilitate analysis and simulation, we propose a deep neural network (DNN) approach to model the antenna radiation pattern functions for different switching states. The proposed DNN achieves a lower normalized mean square error (NMSE) compared to the kernel-based schemes in \cite{kernelfitting2023Gcom,kernelCE2024TC,kernel2010TSP}, with significantly lower computational complexity.

\item  {\bf {Support-Mask-Assisted Sparse Channel Recovery Algorithms with Low Complexity}}:
Achieving low-complexity algorithms for sparse channel estimation involves identifying the superset of the low-dimensional channel support in the angular-delay domain with a high-resolution angular basis. To this end, we propose a modified 2-D orthogonal matching pursuit (OMP) algorithm \cite{OMP} that identifies the shared channel support matrix of vertical and horizontal wireless components. The output from the 2D-OMP algorithm creates a support mask, which is further utilized by a low-complexity variational Bayesian inference (VBI) algorithm \cite{VBI2008,turbo-VBI} to enhance estimation accuracy. We introduce auxiliary variables to develop a modified likelihood model that approximates the 2-D linear minimum mean square error (LMMSE) estimator using two 1-D LMMSE estimators for the sparse transformation from the antenna-frequency domain to the angular-delay domain. Additionally, we propose a Bernoulli-Gaussian prior model to effectively incorporate the support mask, further reducing the complexity of the 1-D LMMSE estimator for sparse channel recovery in the high-resolution angular domain.


%
%


\item  {\bf {Analog Precoding for Discrete FAS State Configuration}}:
Finally, we propose an analog precoding scheme based on the estimated channel. The analog precoder is designed to configure the radiation pattern of each antenna across 12 discrete candidate states. To tackle the combinatorial optimization challenge \cite{greedyoptStatisticCE2023WSA,greedyopt2019TWC}, we reparameterize each discrete state variable as a one-hot state selection vector, which is then approximated as a selection probability vector with a sparse regularization constraint. We apply a gradient-based optimizer \cite{Adam} to find a stationary solution. Simulation results show that our proposed solution significantly outperforms state-of-the-art baselines.

\end{itemize}

\section{System Model}\label{sec:system_model}

\subsection{Brief Introduction on the Pixel-based Fluid Antenna}
The evolution of FAS has been largely influenced by progress in wireless technology, which has produced a variety of outcomes. Despite this, there are relatively few FAS antenna designs, and most of these have been based on mechanical methods, such as liquid-based \cite{liquid1,liquid2}, surface-wave-based \cite{surface-wave1}, and programmable-droplet-based configurations \cite{droplet}. These mechanical antennas can generally meet FAS specifications by manipulating metal or liquid within the antenna to achieve precise spatial sampling. However, because these designs rely on physical movement, the spatial sampling process is relatively slow, which significantly limits FAS performance \cite{s-FAMA}. When compared to the packet transmission rate of about one millisecond per packet, current mechanical antenna designs are not quick enough \cite{BruceLee} to allow for the packet-by-packet reconfiguration that FAS demands.

\begin{figure}[!t]
		\centering
		\includegraphics[width=0.8\linewidth]{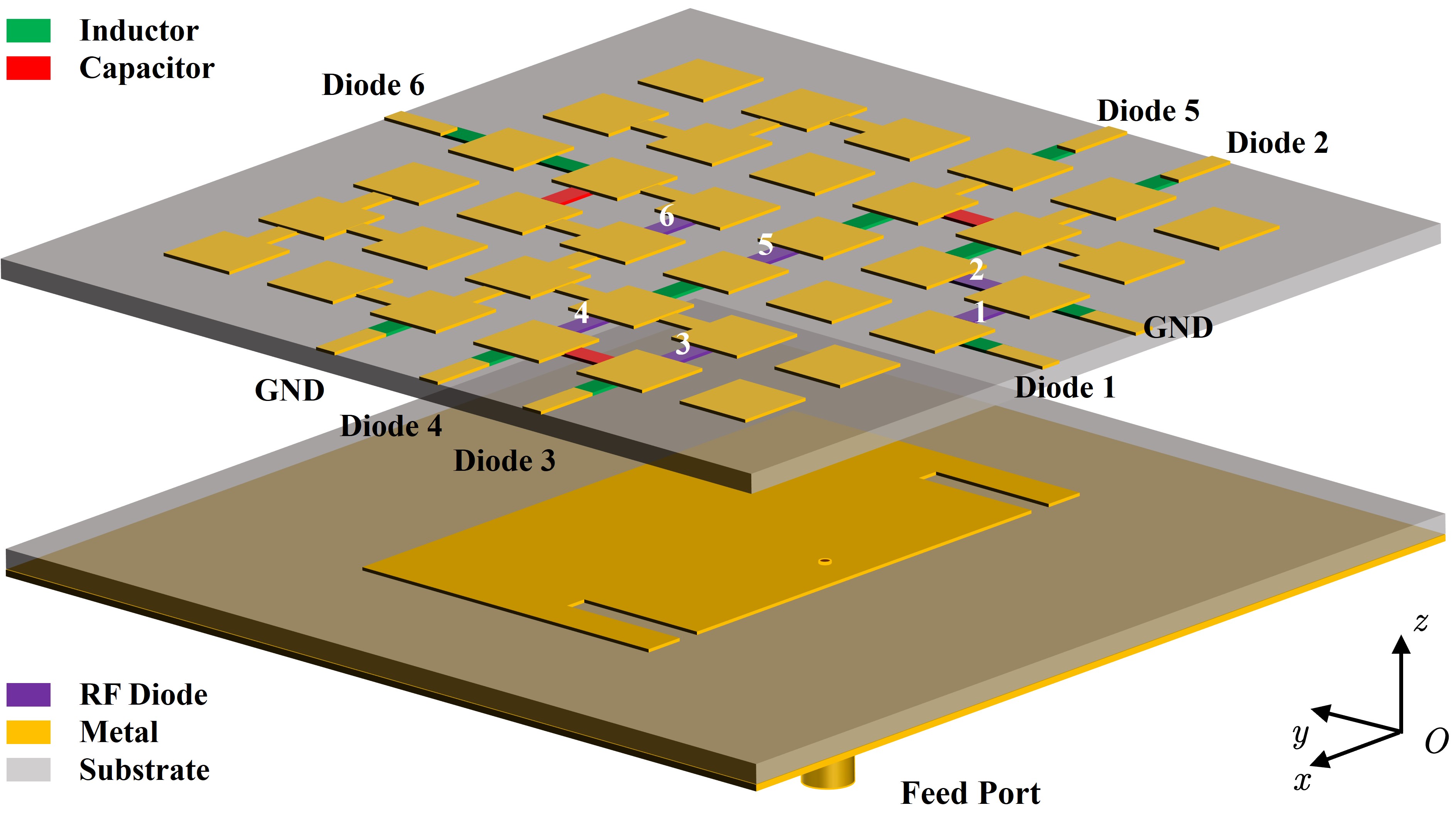}
		\caption{The 3D scheme of the proposed P-FAS in \cite{Pixelantenna2025}.}
		\label{Pixel-FA}
	\end{figure}

\begin{figure}[!t]
		\centering
		\vspace{-0.5cm}
		\subfigure[]{\includegraphics[height=4.3cm]{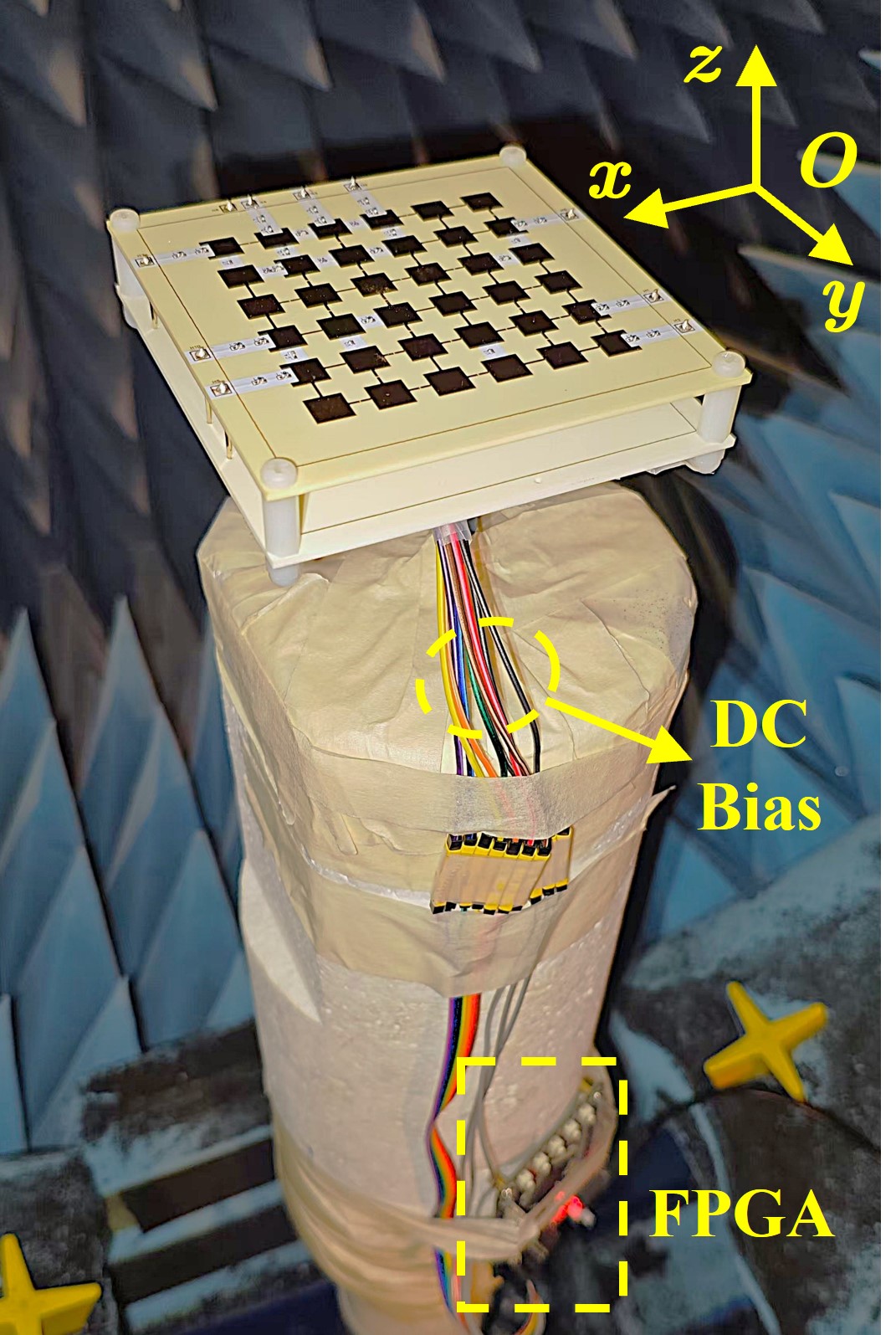}}
		\hfil
		\subfigure[]{\includegraphics[height=4.6cm]{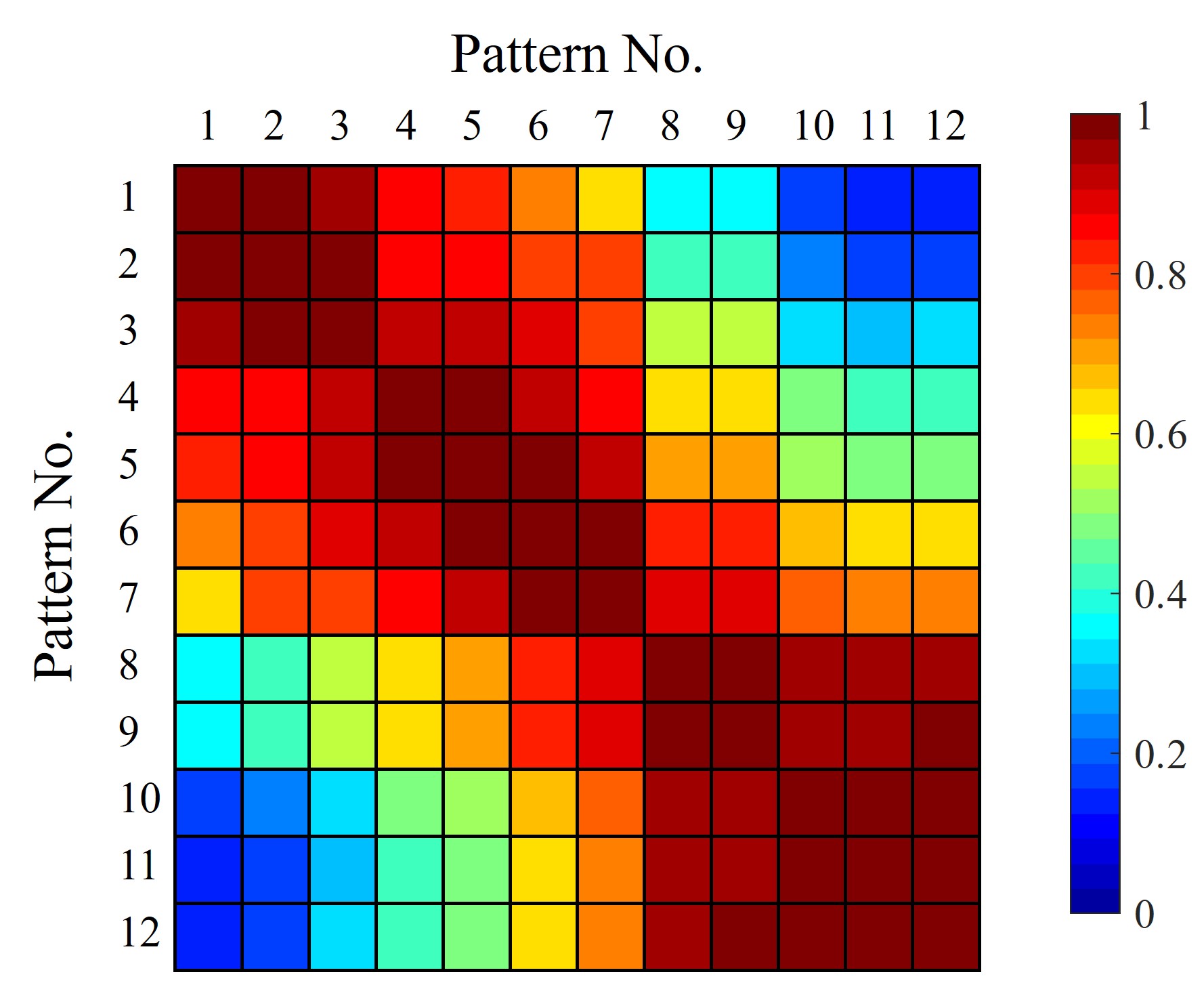}}
		\caption{Illustration of the proposed P-FAS  in \cite{Pixelantenna2025}: (a) Prototype; (b) Pattern correlation matrix.}
		\label{Prototype}
	\end{figure}

The pixel-based fluid antenna represents a transformative advancement in FAS technology, enabling microsecond-level fast reconfigurability through RF switches, such as PIN diodes. This design facilitates rapid antenna pattern switching, which is crucial for meeting the stringent requirements of packet-to-packet adaptability and high spatial resolution in FAS. As illustrated in Figures \ref{Pixel-FA} and \ref{Prototype}(a), the antenna's architecture integrates reconfigurable pixel elements within a compact prototype \cite{Pixelantenna2025}, allowing dynamic adjustment of radiation patterns while maintaining structural simplicity. By leveraging fully scattered electromagnetic environments, the pixel-based FAS mimics the behavior of mechanically fluid antennas (M-FAS), effectively mapping spatial correlation to radiation pattern correlation and replacing physical port switching with reconfigurable state switching.
	
	The key to the pixel-based fluid antenna lies in generating radiation patterns whose covariance matrix aligns with theoretical relationships, such as the Bessel function correlation \cite{FAS2020TWC@KKWong}. As demonstrated in Figure \ref{Prototype}(b) from \cite{Pixelantenna2025}, the pixelated design achieves near-ideal spatial correlation properties, approximating a 12-port M-FAS system across half-wavelength space. This equivalence bridges the gap between spatial diversity and pattern reconfigurability, enabling the antenna to emulate the statistical behavior of a physically movable antenna. Such capabilities are pivotal for FAS applications, where maintaining channel diversity in static environments relies on rapid, precise pattern adjustments rather than mechanical repositioning.
	
	The main advantages of the pixel-based fluid antenna include:
	
	\textit{(a) Fast Switching:} Utilizing RF switches and a software-controllable mechanism, the proposed pixel-based fluid antenna system achieves port switching with microsecond latency, meeting the switching requirements of FAS \cite{f-FAMA,s-FAMA}.
	
	\textit{(b) Easy Control:} The reconfigurability is based on RF switches, which can be easily configured using digital signals. This makes control significantly easier compared to traditional mechanically movable antennas or liquid-based antennas, which require additional motors or pumps.
	
\subsection{Signal Model for the Pixel-based Fluid-Antenna-Assisted Multiuser MIMO-OFDM Systems}
\subsubsection{Channel Model}
We consider a TDD  MISO-OFDM system serving $K$ users, as illustrated in {\figurename~\ref{FAS_illustration}(b)}.
In this system, $M$ pixel-based fluid antennas are deployed at the BS,  while each user is equipped with a single antenna. The system utilizes OFDM with $N_{\rm c}$ subcarriers and operates in a propagation environment characterized by a limited number of scatterers \cite{COST2100}. According to the 3GPP channel model \cite{138901}, the channel impulse response for the $k$-th user at the ${n_{\rm c}}$-th subcarrier is given by
\begin{equation}\label{equ:signal_model}
\begin{aligned}[b]
{\bf h}_{n_{\rm c},k}({\bf s})& =
\sum_{i=1}^I \bigg(
{\rm{diag}} \big(
 {\bm \nu}_{\rm V}(
 {\bm \phi}_k^{(i)} ; {\bf s}
 )  \big)
{\bf h}^{(i)}_{{\rm V},n_{\rm c},k}
 \\
 &+{\rm{diag}} \big(
 {\bm \nu}_{\rm H}(
 {\bm \phi}_k^{(i)} ; {\bf s}
 )  \big)
{\bf h}^{(i)}_{{\rm H},n_{\rm c},k} \bigg)
{e^{ - \jmath \frac{{2\pi {\tau_{i}} \left( {n_{\rm c} - 1} \right)}}{{{N_{\rm{c}}}}}}},
\end{aligned}
\end{equation}
where ${\bf s}=[s_{1}, s_{2}, \cdots, s_{M}]^{\rm T}$ denotes the state configuration vector for the pixel-based FAA, with each antenna element having $L$ candidate states such that $s_m \in \{1,2,\cdots,L\}$. The notation ${\bm \phi}_k^{(i)}=\{ \varphi_k^{(i)}, \theta_k^{(i)}\}$ represents the azimuth and elevation angles for the $i$-th scattering path from the BS array, with ${\tau_{i}}$ indicating the path delay.
The vectors ${\bm \nu}_{\rm V}({\bm \phi}_k^{(i)} ; {\bf s})$ and ${\bm \nu}_{\rm H}({\bm \phi}_k^{(i)} ; {\bf s})$ correspond to the antenna array responses in the vertical and horizontal polarizations, respectively. The
$m$-th elements of these vectors are derived from the reconfigurable antenna radiation patterns ${\nu}_{\rm V}({\bm \phi}_k^{(i)} ; {s}_m)$ and ${\nu}_{\rm H}({\bm \phi}_k^{(i)} ; {s}_m)$. The channel components for the horizontal and vertical polarizations are denoted by ${\bf h}^{(i)}_{{\rm V},n_{\rm c},k}$ and ${\bf h}^{(i)}_{{\rm H},n_{\rm c},k}$, respectively, which are defined by:
\begin{align}
&{\bf h}^{(i)}_{{\rm V},n_{\rm c},k} =
\psi_{{\rm V},k}^{(i)} {\bm \beta}(
 {\bm \phi}_k^{(i)}),\label{equ:hV_model}\\
&{\bf h}^{(i)}_{{\rm H},n_{\rm c},k} =
\psi_{{\rm H},k}^{(i)} {\bm \beta}(
 {\bm \phi}_k^{(i)}), \label{equ:hH_model}
\end{align}
where $\psi_{{\rm V},k}^{(i)} \in {\mathbb C}$ and $\psi_{{\rm H},k}^{(i)} \in {\mathbb C}$ are the response coefficients for the $i$-th path in different polarization directions, and ${\bm \beta}( {\bm \phi}_k^{(i)})$ represents the steering vector. We assume that the BS is equipped with a uniform planar array (UPA) consisting of $M_1$ rows and  $M_2$ columns, such that $M=M_1\times M_2$. The steering vector is then given by:
\begin{align}
{\bm \beta}(
 {\bm \phi})&=\bar{\bf f}_{M_2}(\cos(\theta)) \otimes \bar{\bf f}_{M_1}(\sin(\theta)\cos(\varphi)), \\
\bar{\bf f}_{\bar{M}}(x) &= \frac{1}{\sqrt{\bar{M}}}[1, e^{-\jmath \pi x},\cdots,e^{-\jmath \pi (\bar{M}-1) x}]^{\rm T}.
\end{align}

\begin{figure}
[!t]
\centering
\includegraphics[width=0.8\columnwidth]{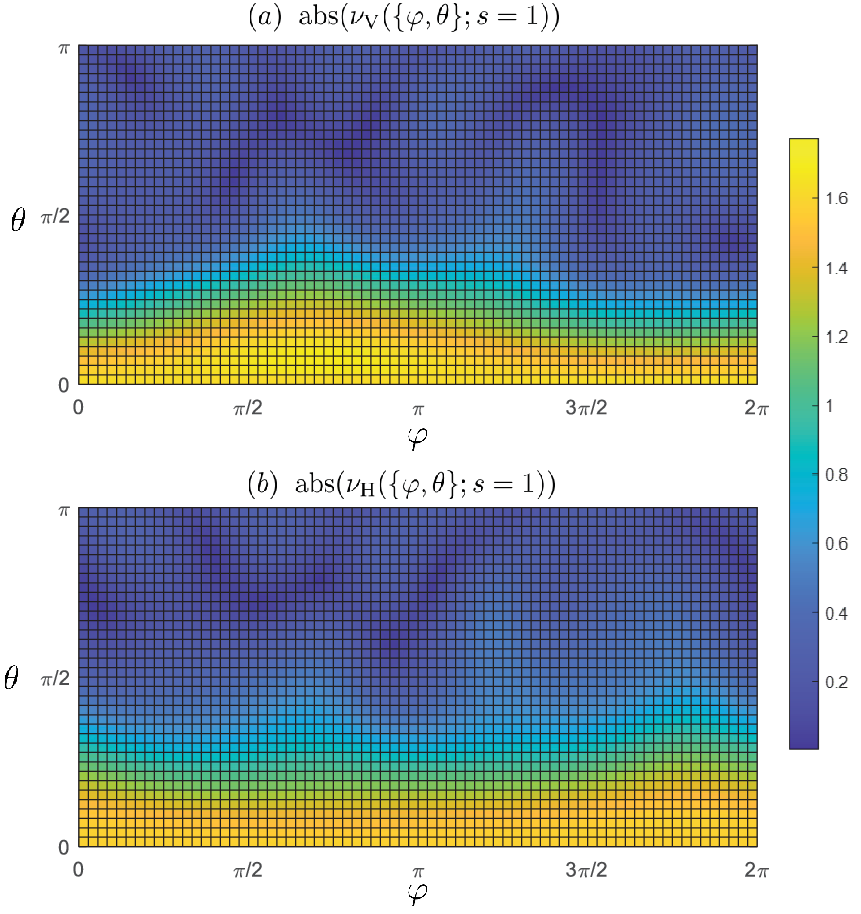}
\caption{Illustration of the magnitudes of ${\nu}_{\rm V}({\bm \phi} ; {s})$ and ${\nu}_{\rm H}({\bm \phi} ; {s})$ for state $s=1$.}
\label{resp_state_1}
\end{figure}

\underline{\emph{\bf State-Non-Separable  Issue}}:
We illustrate the magnitudes of the functions  ${\nu}_{\rm V}({\bm \phi} ; {s})$ and ${\nu}_{\rm H}({\bm \phi} ; {s})$ for state $s=1$ in  {\figurename~\ref{resp_state_1}}. The radiation patterns ${\nu}_{\rm V}({\bm \phi} ; {s})$ and ${\nu}_{\rm H}({\bm \phi} ; {s})$ exhibit intricate dependencies on the angle ${\bm \phi}$. As a result, the array response vectors  ${\bm \nu}_{\rm V}( {\bm \phi}_k^{(i)} ; {\bf s} )$ (or ${\bm \nu}_{\rm H}( {\bm \phi}_k^{(i)} ; {\bf s} )$) and the channel components ${\bf h}^{(i)}_{{\rm V},n_{\rm c},k}$ (or ${\bf h}^{(i)}_{{\rm H},n_{\rm c},k}$) are coupled through the angle ${\bm \phi}_k^{(i)}$. Consequently, the state configuration vector $\bf s$ cannot be treated as a separable term in the expression for  ${\bf h}_{n_{\rm c},k}({\bf s})$, which complicates the channel estimation process.

\subsubsection{Uplink Channel Estimation}
\begin{figure}
[!t]
\centering
\includegraphics[width=.90\columnwidth]{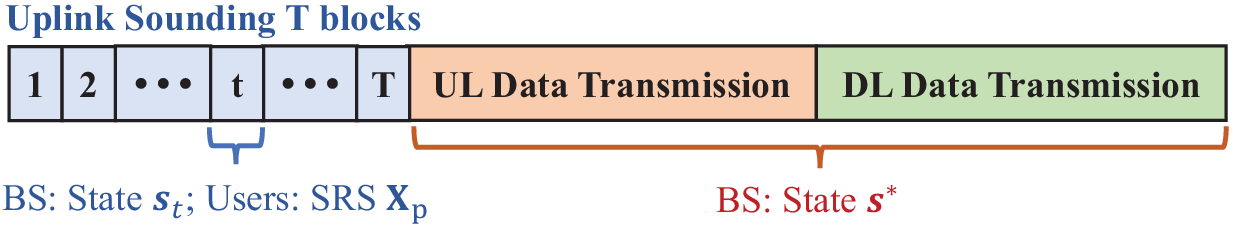}
\caption{Frame structure for the TDD system.}
\label{tdd_frame}
\end{figure}

Due to the state-non-separable issue, the configuration of  $\bf s$ necessitates the estimation of the angles ${\bm \phi}_k^{(i)}$ in order to derive the array responses (${\bm \nu}_{\rm V}$ and ${\bm \nu}_{\rm H}$) as well as the channel components (${\bf h}^{(i)}_{{\rm V},n_{\rm c},k}$ and ${\bf h}^{(i)}_{{\rm H},n_{\rm c},k}$) for all scattering paths. To achieve this goal, we adopt a TDD frame structure, as illustrated in {\figurename~\ref{tdd_frame}}, which consists of $T$ uplink channel sounding blocks. In each block, the BS configures the fluid antenna array to the state ${\bf s}_t$ and triggers uplink sounding reference signals (SRS) ${\bf X}_{\rm p}=[{\bf x}_{{\rm p},1},{\bf x}_{{\rm p},2},\cdots,{\bf x}_{{\rm p},K}]\in {\mathbb C}^{N_{\rm c} \times K}$  from users to the BS. The received signal at the BS for the $t$-th block is given by
\begin{equation}\label{equ:UL_signal_user_all}
\begin{aligned}[b]
{\bf Y}_{n_{\rm c},t}({\bf s}_t)&=\sum_{k=1}^K {\bf h}_{n_{\rm c},k}({\bf s}_t) {x}_{{\rm p},n_{\rm c},k}+ {\bf Z}_{n_{\rm c},t},
\end{aligned}
\end{equation}
where ${\bf Z}_{n_{\rm c},t}$ denotes the additive white Gaussian noise (AWGN), with its elements defined as $z_{n_{\rm c},t,M,k}\sim {\cal{CN}}(0,\sigma_{\rm z}^2)$.

\begin{figure*}
[!t]
\centering
\includegraphics[width=1.6\columnwidth]{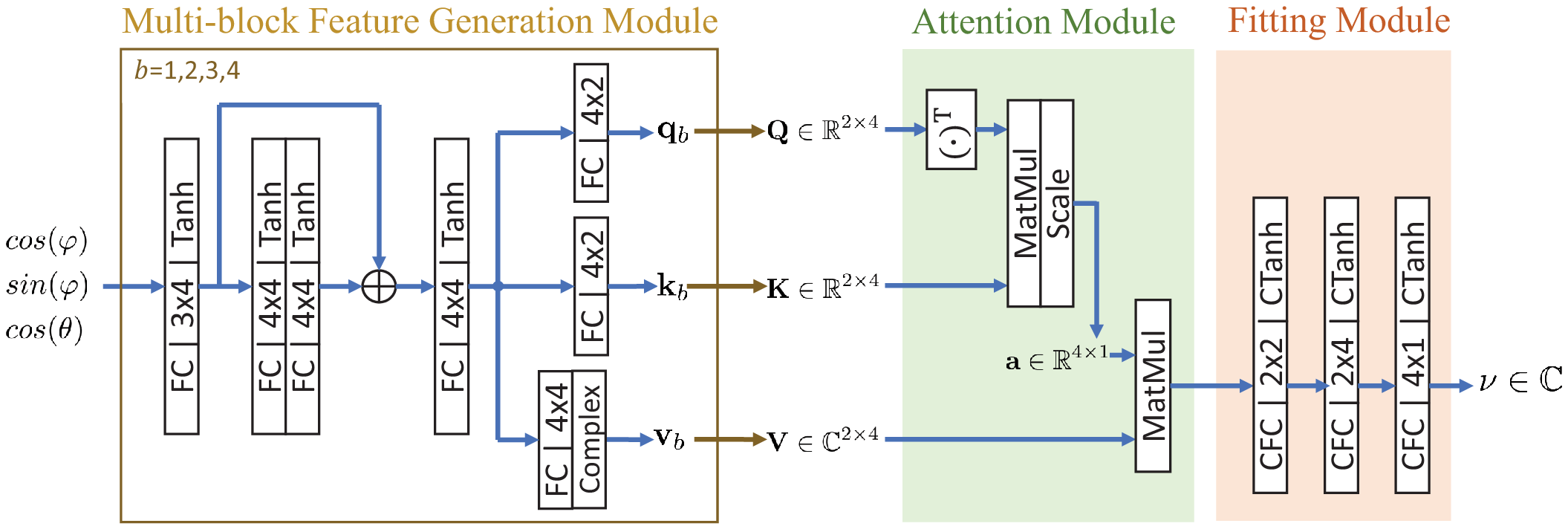}
\caption{Architecture of the proposed DNN. Note that ``FC 4$\times$2'' means a fully-connected layer with 4 input channel and 2 output channel, ``CFC 4$\times$2'' means a complex fully-connected layer, and ``CTanh'' means a complex activation layer using Tanh($\cdot$) function.}
\label{dnn_resp}
\end{figure*}

\subsubsection{Downlink Precoding}
For downlink transmission, the received signal at the $k$-th user is given by
\begin{equation}\label{equ:DL_signal_user_k}
\begin{aligned}[b]
\overline{y}_{n_{\rm c},k}&={\bf h}^{\rm T}_{n_{\rm c},k}({\bf s}^\ast)  {\bf W}_{n_{\rm c}} {\bf x}_{{\rm d}}+ {\overline z}_{n_{\rm c},k},
\end{aligned}
\end{equation}
where  ${\bf x}_{{\rm d}} \in {\mathbb C}^K$ denotes the downlink transmit signal, ${\overline z}_{n_{\rm c},k} \sim {\cal{CN}}(0,\sigma_{\overline{\rm z}}^2) $ represents the AWGN, and ${\bf W}_{n_{\rm c}}$ is the digital precoder. To simplify the analysis and focus on the analog precoder design, we utilize a zero-forcing digital precoder, which is given by
\begin{equation}\label{equ:precoder}
\begin{aligned}[b]
{\bf W}_{n_{\rm c}}&= \sqrt{\gamma_{n_{\rm c}}({\bf s}^\ast)} {\bf H}^\star_{n_{\rm c}}({\bf s}^\ast) \left({\bf H}^{\rm T}_{n_{\rm c}}({\bf s}^\ast) {\bf H}^\star_{n_{\rm c}}({\bf s}^\ast) \right)^{-1},
\end{aligned}
\end{equation}
where ${\bf H}_{n_{\rm c}}({\bf s}^\ast)=[{\bf h}_{n_{\rm c},1}({\bf s}^\ast),{\bf h}_{n_{\rm c},2}({\bf s}^\ast),\cdots,{\bf h}_{n_{\rm c},K}({\bf s}^\ast)] \in {\mathbb C}^{M \times K}$, and $\gamma_{n_{\rm c}}({\bf s}^\ast) \in {\mathbb{R}}^+$ is introduced to ensure that the average transmit power for each subcarrier equals $K P_{\rm T}$, which is given by
\begin{equation}\label{equ:power_zf}
\begin{aligned}[b]
{\gamma_{n_{\rm c}}({\bf s}^\ast)}= \frac{  K  P_{\rm T}} {  {\rm Tr} \left(
\left({\bf H}^{\rm T}_{n_{\rm c}}({\bf s}^\ast) {\bf H}^\star_{n_{\rm c}}({\bf s}^\ast) \right)^{-1}
\right)}.
\end{aligned}
\end{equation}
In this design, the achievable rates for different users and across different subcarriers are the same:
\begin{equation}\label{equ:downlink_rate_2}
{R}_{n_{\rm c},k}({\bf s}^\ast)= \log_2 \left( 1+ \frac{\gamma_{n_{\rm c}}({\bf s}^\ast)}{\sigma_{\overline{\rm z}}^2} \right)
.
\end{equation}
The next task is to perform analog precoding to optimize the state of the pixel-based fluid antenna array $\bf s$:
\begin{align}
    {\mathcal{P}}{(\text{A})}
  \;\;\; {\bf{s}}^\ast  & =\; {\rm{argmax}}_{  \bf{s}
 } \; \;
 \frac{1}{N_{\rm c}}\sum_{n_{\rm c}=1}^{N_{\rm c}} {R}_{n_{\rm c},k}({\bf s}) \notag\\
 & \; {\text{s.t.}} \;  s_m \in \{1,2,\cdots,N_{\rm s}\}, \; \forall  m, \label{equ:discrete_1}
\end{align}
The key issue is the discrete constraint $s_m \in \{1,2,\cdots,N_{\rm s}\}$, which complicates the optimization of ${\mathcal{P}}{(\text{A})}$.

\subsection{DNN-based Modeling on the Antenna Radiation Patterns}
The signal model in \eqref{equ:signal_model} depends on the antenna radiation pattern functions ${\nu}_{\rm V}({\bm \phi} ; {s})$ and ${\nu}_{\rm H}({\bm \phi} ; {s})$, which are intricate functions of ${\bm \phi}=\{ \varphi, \theta\}$, as shown in {\figurename~\ref{resp_state_1}}. Since closed-form expressions for the true functions are difficult to obtain, we employ an approximation method to derive a low-complexity model using data samples from the P-FAS prototype reported in \cite{Pixelantenna2025} to assist algorithm design and performance evaluation.


Although the Legendre-kernel-based method proposed in \cite{kernelfitting2023Gcom,kernelCE2024TC,kernel2010TSP} effectively approximates general antenna radiation patterns, its implementation involves Legendre functions that lead to high computational complexity. To address this challenge, we propose a DNN-based solution with an architecture composed of simple functions. In particular, the proposed DNN architecture is illustrated in {\figurename~\ref{dnn_resp}}. Similar to the Legendre-kernel-based method \cite{kernelfitting2023Gcom,kernelCE2024TC,kernel2010TSP}, the input ${\bm \phi}=\{ \varphi, \theta\}$ is first transformed into $\{\cos(\varphi), \sin(\varphi), \cos(\theta)\}$ before being sent to the feature generation module. This module consists of four parallel blocks that construct features with low complexity, likewise a piecewise function. The generated features are then passed to the attention module to obtain an effective feature vector, which is used by the final fitting module to output the desired antenna radiation pattern response.

\begin{table*}[!t]
\footnotesize
\renewcommand{\arraystretch}{1.1}
\caption{NMSE of the antenna radiation pattern models.}
\label{table_nmse}
\centering
\begin{tabular}{|c|c|c|c|c|c|c|c|c|c|c|c|c|}
\hline
${\nu}_{\rm H}({\bm \phi} ; {s})$     & State 1 & State 2 & State 3 & State 4 & State 5 & State 6 & State 7 & State 8 & State 9 & State 10 & State 11 & State 12
     \\ \hline
Legendre Kernel & 1.4e-3 & 1.0e-3 & 8.4e-4 & 2.9e-4 & 4.5e-4 & 4.6e-4 & 6.8e-4 & 1.3e-3  & 1.3e-3 & 2.0e-3 & 2.5e-3 & 2.5e-3
     \\ \hline
Proposed DNN  & 1.1e-4 & 9.8e-5 & 6.1e-5 & 6.6e-5 & 4.2e-5 & 5.8e-5 & 4.9e-5 & 6.7e-5  & 8.4e-5 & 8.3e-5 & 1.1e-4 & 8.0e-5
     \\ \hline
\hline
${\nu}_{\rm V}({\bm \phi} ; {s})$     & State 1 & State 2 & State 3 & State 4 & State 5 & State 6 & State 7 & State 8 & State 9 & State 10 & State 11 & State 12
     \\ \hline
Legendre Kernel & 1.7e-3 & 1.5e-3 & 1.9e-3 & 1.6e-3 & 2.7e-3 & 3.6e-3 & 4.2e-3 & 4.2e-3  & 4.3e-3 & 3.3e-3 & 3.5e-3 & 3.5e-3
     \\ \hline
Proposed DNN  & 1.1e-4 & 1.4e-4 & 1.7e-4 & 2.8e-4 & 4.5e-4 & 4.3e-4 & 2.9e-4 & 2.3e-4  & 2.1e-4 & 1.8e-4 & 1.4e-4 & 1.8e-4
     \\ \hline
\end{tabular}
\end{table*}

\begin{table}[!t]
\footnotesize
\renewcommand{\arraystretch}{1.1}
\caption{Complexity of the antenna radiation pattern models.}
\label{complexity}
\centering
\begin{tabular}{|c|c|c|}
\hline
         & Legendre Kernel &  Proposed DNN   \\ \hline
FLOPs &  $\geq$1000       & 432          \\ \hline
Parameters &  200       & 510        \\ \hline
\end{tabular}
\\
\vspace{1ex}
\scriptsize
{\raggedright {\emph{Notes}:} The use of ``$\geq$'' is due to the fact that the complexity of the Legendre function has not yet been accounted for. \par}
\end{table}

We compare the modeling accuracy and complexity of the proposed DNN with the Legendre-kernel-based method, as shown in Table \ref{table_nmse} and Table \ref{complexity}. Both models are trained on $10^7$ data samples and tested on a dataset of $10^6$ samples, using the Adam optimizer \cite{Adam} in PyTorch \cite{NEURIPS2019_9015} with a learning rate of $10^{-4}$ and a maximum of $400$ epochs. The results show that the proposed DNN reduces the NMSE by more than a factor of 10 in most cases, while also achieving significantly lower computational complexity.

\section{Modified Orthogonal Matching Pursuit for Channel Estimation}
In this section, we first introduce an approximate separable channel response model to facilitate channel estimation. Building on this proposed channel model, we present a sparse channel recovery algorithm designed to jointly estimate the angles, delays, and channel components for all scattering rays, while ensuring low computational complexity.

\subsection{Approximate Separable Channel Response Model}
To address the state-non-separable issue, we consider an approximate model in which the delay $0 \leq {\tau} \leq (L-1)$ is an integer, and the angle ${\bm \phi} \in {\mathcal B}$, where ${\mathcal B}$ is a grid-based angle set created by uniformly sampling angles from $0\leq \theta \leq \pi$ and $0\leq \varphi < 2\pi$. Consequently, the channel for the $k$-th user at the ${n_{\rm c}}$-th subcarrier, as shown in \eqref{equ:signal_model},  can be approximately expressed as
\begin{equation}\label{equ:signal_model_app}
\begin{aligned}[b]
{\bf h}_{n_{\rm c},k}({\bf s})& =
\sum_{b=1}^{|{\mathcal B}|} \sum_{\ell=0}^{L-1}
\bigg( {\rm{diag}} \big(
 {\bm \nu}_{\rm V}(
 {\bm \phi}^{(b)} ; {\bf s}
 )  \big) {\bm \beta}(
 {\bm \phi}^{(b)})\psi_{{\rm V},k}^{(b,\ell)} \\
 & \;
  + {\rm{diag}} \big(
 {\bm \nu}_{\rm H} (
 {\bm \phi}^{(b)} ; {\bf s}
 )  \big) {\bm \beta}(
 {\bm \phi}^{(b)})
\psi_{{\rm H},k}^{(b,\ell)}  \bigg)
{e^{ - \jmath \frac{{2\pi {\tau_{\ell}} \left( {n_{\rm c} - 1} \right)}}{{{N_{\rm{c}}}}}}},
\end{aligned}
\end{equation}
where ${\bm \phi}^{(b)}$ is the $b$-th element in ${\mathcal B}$.
It is seen from equation \eqref{equ:signal_model_app}, channel estimation now involves estimating the response coefficients $\psi_{{\rm V},k}^{(b,\ell)}$ and $\psi_{{\rm H},k}^{(b,\ell)}$ for all $\ell$ and $b$, which are the separable terms in the channel model.

\begin{figure}
[!t]
\centering
\includegraphics[width=.80\columnwidth]{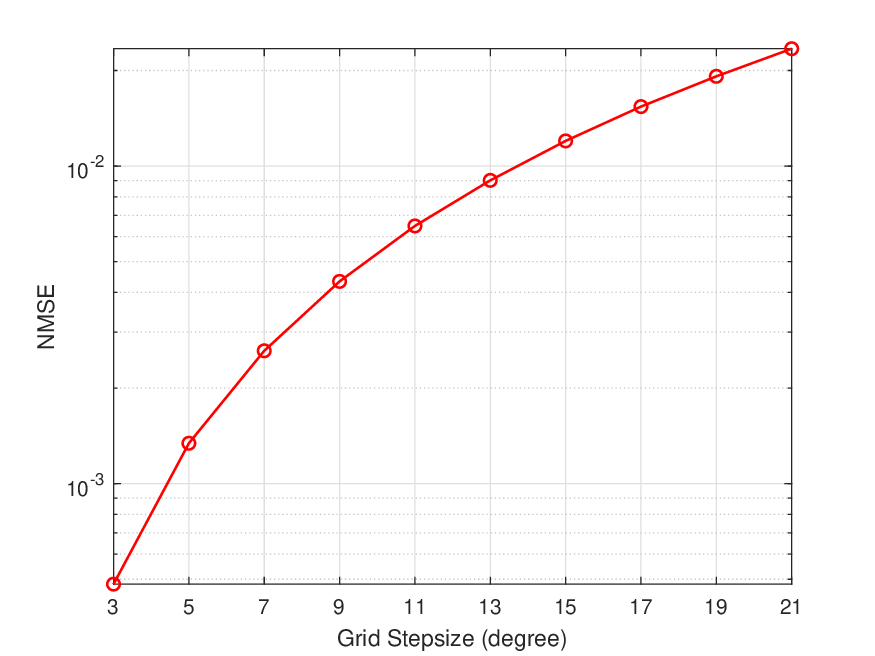}
\caption{Illustration of the approximation accuracy on the antenna radiation patterns for the grid-based angle set with different grid stepsizes.}
\label{grid_resolution}
\end{figure}

The approximation loss in this model arises from replacing the exact antenna radiation patterns ${\bm \nu}_{\rm V}( {\bm \phi}_k^{(i)} ; {\bf s} )$ and ${\bm \nu}_{\rm H}( {\bm \phi}_k^{(i)} ; {\bf s} )$ with approximated values related to the angles on the grid. We further assess the effectiveness of this approximation using the grid-based angle set shown in {\figurename~\ref{grid_resolution}}. The results show that the approximation accuracy can reach nearly $10^{-3}$ when the angle grid resolution is set to $5^\circ$, which includes 2701 complex elements.

Define  ${\bf G}_{\rm V}({\bf s})=[{\bf g}_{{\rm V},1}({\bf s}),{\bf g}_{{\rm V},2}({\bf s}),\cdots,{\bf g}_{{\rm V},{|{\mathcal B}|}}({\bf s})]\in {\mathbb C}^{M \times |{\mathcal B}|}$ and ${\bf G}_{\rm H}({\bf s})=[{\bf g}_{{\rm H},1}({\bf s}),{\bf g}_{{\rm H},2}({\bf s}),\cdots,{\bf g}_{{\rm H},{|{\mathcal B}|}}({\bf s})]\in {\mathbb C}^{M \times |{\mathcal B}|}$ as the bases for the angular domain given the state vector $\bf s$, where
\begin{align}
{\bf g}_{{\rm V},b}({\bf s})&=
{\rm{diag}} \big(
 {\bm \nu}_{\rm V}(
 {\bm \phi}^{(b)} ; {\bf s}
 )  \big) {\bm \beta}(
 {\bm \phi}^{(b)}), \\
 {\bf g}_{{\rm H},b}({\bf s})&=
{\rm{diag}} \big(
 {\bm \nu}_{\rm H}(
 {\bm \phi}^{(b)} ; {\bf s}
 )  \big) {\bm \beta}(
 {\bm \phi}^{(b)}).
\end{align}
Letting ${\bf H}_{k}({\bf s})=[{\bf h}_{1,k}({\bf s}),{\bf h}_{2,k}({\bf s}),\cdots,{\bf h}_{N_{\rm c},k}({\bf s})]$, we have
\begin{equation}\label{equ:Hk}
\begin{aligned}[b]
{\bf H}_{k}({\bf s})&=\left( {\bf G}_{\rm V}({\bf s})  {\bm \Psi}_{{\rm V},k} +{\bf G}_{\rm H}({\bf s}) {\bm \Psi}_{{\rm H},k} \right) {\bf F}^{\rm T},
\end{aligned}
\end{equation}
where ${\bf F} \in {\mathbb C}^{N_{\rm c} \times L}$ is the partial discrete Fourier transform (DFT) matrix, and ${\bm \Psi}_{{\rm V},k} \in {\mathbb C}^{|{\mathcal B}| \times L}$ and ${\bm \Psi}_{{\rm H},k} \in {\mathbb C}^{|{\mathcal B}| \times L}$ represent the matrix forms of $\psi_{{\rm V},k}^{(b,\ell)}$ and $\psi_{{\rm H},k}^{(b,\ell)}$ for all $b$ and $\ell$, respectively. Finally, let $\widetilde{\bf G}({\bf s})=[{\bf G}_{\rm V}({\bf s}),{\bf G}_{\rm H}({\bf s})]$ and $\widetilde{\bm \Psi}_{k}=[{\bm \Psi}_{{\rm V},k}^{\rm T},{\bm \Psi}_{{\rm H},k}^{\rm T}]^{\rm T}$. The channel is then represented by
\begin{equation}\label{equ:Hk2}
\begin{aligned}[b]
{\bf H}_{k}({\bf s})&=\widetilde{\bf G}({\bf s}) \widetilde{\bm \Psi}_{k} {\bf F}^{\rm T}.
\end{aligned}
\end{equation}

\subsection{Estimation Problem Formulation}

Assuming orthogonal uplink SRS pilots, channel estimation for different users can be performed independently. For simplicity, we will omit the index, and the signal model is as follows:
\begin{equation}\label{equ:UL_signal_user_k_v2}
\begin{aligned}[b]
\widetilde{\bf y}_{n_{\rm c},t}({\bf s}_t)
&= {\bf h}_{n_{\rm c}}({\bf s}_t) {x}_{{\rm p},n_{\rm c}} + \widetilde{\bf z}_{n_{\rm c},t},
\end{aligned}
\end{equation}
where $\widetilde{\bf y}_{n_{\rm c},t}({\bf s}_t)$ and $\widetilde{\bf z}_{n_{\rm c},t}$ are the received signal and noise after dropping the pilots.
We further define $\widetilde{\bf Y}_{t}({\bf s}_t)=[\widetilde{\bf y}_{1,t}({\bf s}_t),\cdots,\widetilde{\bf y}_{n_{\rm c},t}({\bf s}_t)]$ and $\widetilde{\bf Z}_{t}=[\widetilde{\bf z}_{1,t},\cdots,\widetilde{\bf z}_{n_{\rm c},t}]$, and then we have
\begin{equation}\label{equ:UL_signal_Yt}
\begin{aligned}[b]
\widetilde{\bf Y}_{t}({\bf s}_t)
&= {\bf H}({\bf s}_t) {\rm diag}({\bf x}_{\rm p}) + \widetilde{\bf Z}_{t}\\
&=\widetilde{\bf G}({\bf s}_t) \widetilde{\bm \Psi} \overline{\bf F}^{\rm T}+ \widetilde{\bf Z}_{t},
\end{aligned}
\end{equation}
where $\overline{\bf F}={\rm diag}({\bf x}_{\rm p}){\bf F}$ denotes the pilot-weighted partial DFT.

We define ${\bf G}_{\rm uv}=[{\bf G}_{\rm V}({\bf s}_1)^{\rm T},\cdots,{\bf G}_{\rm V}({\bf s}_T)^{\rm T}]^{\rm T}$ and ${\bf G}_{\rm uh}=[{\bf G}_{\rm H}({\bf s}_1)^{\rm T},\cdots,{\bf G}_{\rm H}({\bf s}_T)^{\rm T}]^{\rm T}$ to summarize the angular bases for all involved states. Letting $\widetilde{\bf G}_{\rm u}=[{\bf G}_{\rm uv},{\bf G}_{\rm uh}]$,
 ${\bf Z}_{\rm u}=[\widetilde{\bf Z}_1^{\rm T},\widetilde{\bf Z}_2^{\rm T},\cdots,\widetilde{\bf Z}_T^{\rm T}]^{\rm T}$, and ${\bf Y}_{\rm u}=[\widetilde{\bf Y}_{1}({\bf s}_1)^{\rm T},\widetilde{\bf Y}_{2}({\bf s}_2)^{\rm T},\cdots, \widetilde{\bf Y}_{T}({\bf s}_T)^{\rm T}]^{\rm T}$, the overall received pilot signal for all the $T$ blocks is given by
\begin{equation}\label{equ:UL_signal_Yu}
\begin{aligned}[b]
{\bf Y}_{\rm u}
&=\widetilde{\bf G}_{\rm u} \widetilde{\bm \Psi} \overline{\bf F}^{\rm T}+ {\bf Z}_{\rm u},
\end{aligned}
\end{equation}
and we further write it by a linear observation model:
\begin{equation}\label{equ:UL_signal_Yu_linear}
\begin{aligned}[b]
{\rm vec}({\bf Y}_{\rm u})
&=\overline{\bf F} \otimes \widetilde{\bf G}_{\rm u} {\rm vec} (\widetilde{\bm \Psi}) + {\rm vec}( {\bf Z}_{\rm u}).
\end{aligned}
\end{equation}

\begin{figure}
[!t]
\centering
\includegraphics[width=.90\columnwidth]{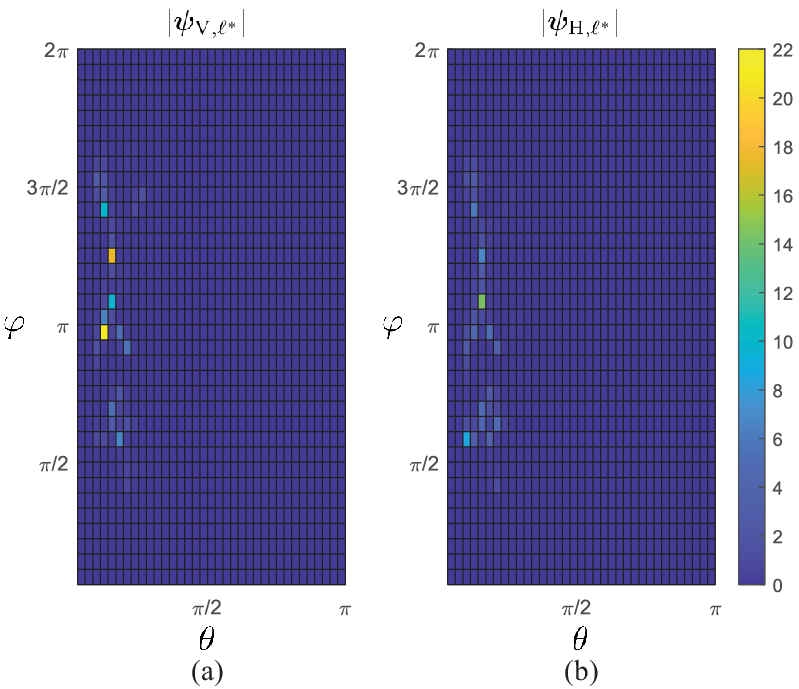}
\caption{Illustration of the elements in ${\bm \Psi}_{{\rm V},\ell^\ast}$ and ${\bm \Psi}_{{\rm H},\ell^\ast}$ after reshaping to the 2-D angular domain, where $\ell^\ast$ is the index of the strongest delay tap.}
\label{angular_resp}
\end{figure}

We  illustrate the elements in the $\ell^\ast$ column of ${\bm \Psi}_{\rm V}$ and ${\bm \Psi}_{\rm H}$ in {\figurename~\ref{angular_resp}}, where $\ell^\ast$ represents the index of the strongest delay tap. The results show that the channel responses ${\bm \psi}_{{\rm V},\ell^\ast}$ and ${\bm \psi}_{{\rm H},\ell^\ast}$ are very sparse in the angular domain, containing only a few non-zero elements, and their supports (i.e., the indices of the non-zero elements) are nearly identical. Therefore, despite the large dimension of $\widetilde{\bm \Psi}$, a sparse recovery algorithm can be used to estimate $\widetilde{\bm \Psi}$ with a small pilot overhead $T$. Specifically, the estimation problem can be formulated as follows:
\begin{align}
    {\mathcal{P}}{(\text{B})}
  \;\;\; \widetilde{\bm \Psi}  & =\; {\rm{argmin}}_{  \widetilde{\bm \Psi}
 } \; \;
  \big\| \left|{\bm \Psi}_{\rm V} \right|+\left|{\bm \Psi}_{\rm H}\right| \big\|_0
 \notag\\
 & \; {\text{s.t.}} \;  \|{\bf Y}_{\rm u}-\widetilde{\bf G}_{\rm u} \widetilde{\bm \Psi} \overline{\bf F}^{\rm T}\|_{\rm F}^2 \leq M N_{\rm c}\sigma_{\rm z}^2, \label{equ:constraint_Pb}
\end{align}
where the objective function indicates that ${\bm \Psi}_{\rm V}$ and ${\bm \Psi}_{\rm H}$ share the same support. Such sharing support feature can be seen from the models in \eqref{equ:hV_model} and \eqref{equ:hH_model}, in which the elements $\psi_{{\rm V}}^{(i)}$ and $\psi_{{\rm H}}^{(i)}$ are the response coefficients for the same $i$-th scattering path but in different polarization directions.

However, designing an effective sparse channel recovery algorithm for ${\mathcal{P}}{(\text{B})}$ presents several challenges:
\begin{itemize}
\item First, the antenna-angular transformation matrix $\widetilde{\bf G}_{\rm u}$ is highly singular for a high-resolution angular grid set ${\mathcal B}$, which limits the effectiveness of most AMP-based algorithms \cite{AMP,VAMP,OAMP,TurboOMP2015SPL}.
\item Second, most sparse recovery algorithms rely on the linear observation model \eqref{equ:UL_signal_Yu_linear}. However, the large dimension of the measurement matrix $\overline{\bf F} \otimes \widetilde{\bf G}_{\rm u}$ makes a low-complexity algorithm more desirable.
\item Third, it is necessary to prune the zero-power rays in the approximated channel model \eqref{equ:signal_model_app} to reduce the complexity of the optimization algorithm for analog precoding.
\end{itemize}

\subsection{Pre-Processing on ${\bf Y}_{\rm u}$ and Least Squares Estimator}\label{sec:Pre-Processing}
To facilitate numerical operations, we  perform signal pre-processing on ${\bf Y}_{\rm u}$ to ensure that the sensing matrix has linearly independent rows, making the right inverse feasible.
Specifically, we apply singular value decomposition (SVD) to $\widetilde{\bf G}_{\rm u}$, yielding $\widetilde{\bf G}_{\rm u} = {\bf U}_{\rm u} {\rm{diag}}({\bf s}_{\rm u}) {\bf V}_{\rm u}^{\rm H}$. We define $\widetilde{\bf U}_{\rm u}$ as the matrix of left-singular vectors corresponding to the first $\widetilde{M}$ dominant singular values in ${\bf s}_{\rm u}$. The new signal model is obtained by transforming ${\bf Y}_{\rm u}$ in \eqref{equ:UL_signal_Yu} into $\widetilde{\bf Y}_{\rm u} = \widetilde{\bf U}_{\rm u}^{\rm H} {\bf Y}_{\rm u}$:
\begin{equation}\label{equ:UL_signal_Yu_svd}
\begin{aligned}[b]
\widetilde{\bf Y}_{\rm u}
&=\widetilde{\bf U}_{\rm u}^{\rm H} {\bf Y}_{\rm u}\\
&=\widetilde{\bf U}_{\rm u}^{\rm H}\widetilde{\bf G}_{\rm u} \widetilde{\bm \Psi} \overline{\bf F}^{\rm T}+ \widetilde{\bf Z}_{\rm u},
\end{aligned}
\end{equation}
where $\widetilde{\bf Z}_{\rm u}={\bf U}_{\rm u}^{\rm H} {\bf Z}_{\rm u}$.
Substituting $\widetilde{\bf G}_{\rm u}=[{\bf G}_{\rm uv},{\bf G}_{\rm uh}]$ and $\widetilde{\bm \Psi}=[{\bm \Psi}_{{\rm V}}^{\rm T},{\bm \Psi}_{{\rm H}}^{\rm T}]^{\rm T}$ into \eqref{equ:UL_signal_Yu_svd}, the observation model is given by
\begin{equation}\label{equ:UL_signal_Yu_svd_2}
\begin{aligned}[b]
\widetilde{\bf Y}_{\rm u}
=\widetilde{\bf U}_{\rm u}^{\rm H}{\bf G}_{\rm uv} {\bm \Psi}_{\rm V} \overline{\bf F}^{\rm T}
+\widetilde{\bf U}_{\rm u}^{\rm H}{\bf G}_{\rm uh} {\bm \Psi}_{\rm H} \overline{\bf F}^{\rm T}+\widetilde{\bf Z}_{\rm u}.
\end{aligned}
\end{equation}
Defining ${\bf Q}_{\rm V}=\overline{\bf F} \otimes (\widetilde{\bf U}_{\rm u}^{\rm H} {\bf G}_{\rm uv})$ and ${\bf Q}_{\rm H}=\overline{\bf F} \otimes (\widetilde{\bf U}_{\rm u}^{\rm H} {\bf G}_{\rm uh})$,  we finally obtain a typical linear observation model:
\begin{equation}\label{equ:UL_signal_Yu_vec}
\begin{aligned}[b]
{\rm{vec}}(\widetilde{\bf Y}_{\rm u})
&={\rm{vec}}(\widetilde{\bf U}_{\rm u}^{\rm H}{\bf G}_{\rm uv} {\bm \Psi}_{\rm V} \overline{\bf F}^{\rm T})\\
&\qquad+{\rm{vec}}(\widetilde{\bf U}_{\rm u}^{\rm H}{\bf G}_{\rm uh} {\bm \Psi}_{\rm H} \overline{\bf F}^{\rm T})+{\rm{vec}}(\widetilde{\bf Z}_{\rm u})\\
&={\bf Q}_{\rm V} {\rm{vec}} ({\bm \Psi}_{\rm V})+{\bf Q}_{\rm H} {\rm{vec}} ({\bm \Psi}_{\rm H})+{\rm{vec}}(\widetilde{\bf Z}_{\rm u}).
\end{aligned}
\end{equation}

Letting $\widetilde{\bf Q}=[{\bf Q}_{\rm V},{\bf Q}_{\rm H}]$, one can estimate ${\bm \Psi}_{\rm V}$ and ${\bm \Psi}_{\rm H}$ using the least squares (LS) method with the right inverse of $\widetilde{\bf Q}$:
\begin{equation}\label{equ:LS_est}
\begin{aligned}[b]
\left[\begin{matrix}
    {\rm{vec}} ({\bm \Psi}_{\rm V})\\
    {\rm{vec}} ({\bm \Psi}_{\rm H})
\end{matrix}\right]
    = \widetilde{\bf Q}^{\rm H} \left(\widetilde{\bf Q} \widetilde{\bf Q}^{\rm H}\right)^{-1}{\rm{vec}}(\widetilde{\bf Y}_{\rm u}).
\end{aligned}
\end{equation}
However, the LS method generally performs poorly since the number of rows in $\widetilde{\bf Y}_{\rm u}$ (which is $\widetilde{M}$) is much smaller than that of ${\bm \Psi}_{\rm V}$ or ${\bm \Psi}_{\rm H}$ (which is $|{\cal B}|L$). To enhance estimation accuracy, a sparse channel recovery algorithm will be proposed in the following subsection.

\subsection{Proposed Modified OMP-Based Estimator}
In this subsection, we propose a low-complexity solution for ${\mathcal{P}}{(\text{B})}$ based on the OMP algorithm to estimate ${\bm \Psi}_{\rm V}$ and ${\bm \Psi}_{\rm H}$ from $\widetilde{\bf Y}_{\rm u}$. To leverage the shared support feature of ${\bm \Psi}_{\rm V}$ and ${\bm \Psi}_{\rm H}$, modifications to the OMP algorithm are necessary.  Before delving into the details of the algorithm design, we first define $\cal D$ as the set of shared support for $ {\bm \Psi}_{\rm V}$ and ${\bm \Psi}_{\rm H}$.\footnote{It is important to clarify that the definition of
$\cal D$ is not rigid. In this subsection, it refers to the support of the vectorized forms of $ {\bm \Psi}_{\rm V}$ and ${\bm \Psi}_{\rm H}$, whereas in the next section, it refers to the support of their original matrix forms. }
The effective elements in ${\bm \Psi}_{\rm V}$ and ${\bm \Psi}_{\rm H}$ are denoted by ${\bm \psi}_{\rm V}^\ast({\cal D})$ and ${\bm \psi}_{\rm H}^\ast({\cal D})$, respectively.  Next, we propose an estimation algorithm, akin to OMP \cite{OMP}, to jointly estimate the set $\cal D$ and the effective channel vectors ${\bm \psi}_{\rm V}^\ast({\cal D})$ and ${\bm \psi}_{\rm H}^\ast({\cal D})$. The overall algorithm is summarized in Algorithm \ref{alg:MOMP}. Specifically, the algorithm comprises two key steps.

\begin{itemize}
\item
{\bf Effective Elements Estimation}:
Given support $\cal D$, the effective measurement matrices can be obtained by selecting the corresponding $|{\cal D}|$ columns from ${\bf Q}_{\rm V}$ and ${\bf Q}_{\rm H}$ according to ${\cal D}$, which are denoted by ${\bf Q}_{\rm V}^\ast({\cal D})$ and ${\bf Q}_{\rm H}^\ast({\cal D})$, respectively. Defining $\widetilde{\bf Q}^\ast({\cal D})=[{\bf Q}_{\rm V}^\ast({\cal D}),{\bf Q}_{\rm H}^\ast({\cal D})]$,
    we can estimate ${\bm \psi}_{\rm V}^\ast({\cal D})$ and ${\bm \psi}_{\rm H}^\ast({\cal D})$ by
    \begin{equation}\label{equ:MOMP_est}
    \begin{aligned}[b]
    \left[\begin{matrix}
    {\bm \psi}_{\rm V}^\ast({\cal D})\\
    {\bm \psi}_{\rm H}^\ast({\cal D})
    \end{matrix}\right]
    =\left(\widetilde{\bf Q}^\ast({\cal D})^{\rm H} \widetilde{\bf Q}^\ast({\cal D}) \right)^{-1} \widetilde{\bf Q}^\ast({\cal D})^{\rm H}{\rm{vec}}(\widetilde{\bf Y}_{\rm u}).
    \end{aligned}
    \end{equation}
    The residual signal is given by
    \begin{equation}\label{equ:MOMP_residual}
    \begin{aligned}[b]
    {\bf r}_{\rm u} &={\rm{vec}}(\widetilde{\bf Y}_{\rm u})-{\bf Q}_{\rm V}^\ast({\cal D}){\bm \psi}_{\rm V}^\ast({\cal D})-{\bf Q}_{\rm H}^\ast({\cal D}){\bm \psi}_{\rm H}^\ast({\cal D}).
    \end{aligned}
    \end{equation}
\item {\bf Support Updating}:
Let the $i$-th column of ${\bf Q}_{\rm V}$ and ${\bf Q}_{\rm H}$ be denoted by ${\bf q}_{{\rm V},i}$ and ${\bf q}_{{\rm H},i}$, respectively. Defining $\widetilde{\bf q}_i=[{\bf q}_{{\rm V},i},{\bf q}_{{\rm H},i}]$, we select the best support index  by
    \begin{equation}\label{equ:MOMP_support}
    \begin{aligned}[b]
    i^\ast = \underset{i \in {\cal D}^{\rm c}}{\rm{argmax}} \; \| (\widetilde{\bf q}_i^{\rm H} \widetilde{\bf q}_i)^{-1}  \widetilde{\bf q}_i^{\rm H} {\bf r}_{\rm u} \|_2^2,
    \end{aligned}
    \end{equation}
    where ${\cal D}^{\rm c}$ is the complement of ${\cal D}$. The support ${\cal D}$ is then updated by ${\cal D}={\cal D} \cup \{i^\ast\}$.
\end{itemize}

\underline{\emph{Complexity Analyses}}: The complexity of the modified OMP in Algorithm \ref{alg:MOMP} is ${\cal O}\left(\widetilde{M}N_{\rm c} |{\cal B}|L+ \widetilde{M}N_{\rm c} N_{\rm a}^2+N_{\rm a}^3\right)$, where $N_{\rm a}$ represents the length of the active support for non-zero elements.

\begin{algorithm}[!t]
\caption{Modified OMP for ${\mathcal{P}}{(\text{B})}$}\label{alg:MOMP}
\begin{algorithmic}[1]
\State\textbf{Input}:  Observation  ${\rm{vec}}(\widetilde{\bf Y}_{\rm u})$ in \eqref{equ:UL_signal_Yu_svd};
\State \textbf{Initialize}: ${\bf r}={\rm{vec}}(\widetilde{\bf Y}_{\rm u})$ and ${\cal D}=\varnothing$;
\While{$\| {\bf r} \|_2^2 > \widetilde{M} N_{\rm c}\sigma_{\rm z}^2 $}
\State Calculate $i^\ast = \underset{i \in {\cal D}^{\rm c}}{\rm{argmax}} \; \| (\widetilde{\bf q}_i^{\rm H} \widetilde{\bf q}_i)^{-1}  \widetilde{\bf q}_i^{\rm H} {\bf r}_{\rm u} \|_2^2$;
\State Update support ${\cal D}={\cal D} \cup \{i^\ast\}$;
\State Estimate ${\bm \psi}_{\rm V}^\ast({\cal D})$ and ${\bm \psi}_{\rm H}^\ast({\cal D})$ using \eqref{equ:MOMP_est};
\State Update residual ${\bf r}_{\rm u}$ using \eqref{equ:MOMP_residual};
\EndWhile
\State Map ${\bm \psi}_{\rm V}^\ast({\cal D})$ and ${\bm \psi}_{\rm H}^\ast({\cal D})$ into ${\bm \Psi}_{\rm V}$ and ${\bm \Psi}_{\rm H}$ according to the support $\cal D$, setting the remaining elements to zero.
\State\textbf{Output}: The estimated ${\bm \Psi}_{\rm V}$ and ${\bm \Psi}_{\rm H}$, and support ${\cal D}$.
\end{algorithmic}
\end{algorithm}

\section{Low-Complexity Bayesian Estimator for Performance Enhancement}
As shown in {\figurename~\ref{angular_resp}}, the magnitudes of the non-zero elements in ${\bm \Psi}_{\rm V}$ and ${\bm \Psi}_{\rm H}$ vary significantly. This variation renders the modified OMP estimator  suboptimal, as this characteristic is not effectively utilized. In this section, we improve estimation accuracy by proposing a Bayesian estimator to learn the magnitudes of the elements, which are then applied in the denoising process with the LMMSE estimator.

Existing Bayesian methods \cite{VAMP,VBI2008,turbo-VBI} encounter complexity issues when directly applied to solve ${\mathcal{P}}{(\text{B})}$ due to the high dimensions of the observation $\widetilde{\bf Y}_{\rm u} \in {\mathbb C}^{\widetilde{M} \times N_{\rm c}}$ and the variables ${\bm \Psi}_{{\rm V}} \in {\mathbb C}^{|{\mathcal B}| \times L}$ and ${\bm \Psi}_{{\rm H}} \in {\mathbb C}^{|{\mathcal B}| \times L}$. To address this issue, we propose an approximate Bayesian inference algorithm that reduces computational complexity through two specific designs:
\begin{itemize}
\item {\bf Approximate 2D-LMMSE by Two 1D-LMMSE}: The LMMSE estimator is commonly used for denoising in existing Bayesian approaches \cite{VAMP,VBI2008,turbo-VBI}, but its complexity depends on the dimension of the observation ${\rm{vec}}(\widetilde{\bf Y}_{\rm u})$. We approximate the 2D-LMMSE using two 1D-LMMSE estimators by applying the approximate message passing technique, significantly reducing overall complexity.
\item {\bf Prior Mask to Reduce the Dimension of Estimated Variables}: Learning the magnitudes of variables in ${\bm \Psi}_{{\rm V}}$ and ${\bm \Psi}_{{\rm H}}$ can be challenging, especially when both $|{\mathcal B}|$ and $L$ are large. However, these variables are highly sparse, meaning only a few non-zero elements need to be estimated. By introducing a prior mask based on the estimated support ${\cal D}$ from the modified OMP in Algorithm \ref{alg:MOMP}, we can significantly reduce the number of effective elements to be estimated in ${\bm \Psi}_{{\rm V}}$ and ${\bm \Psi}_{{\rm H}}$.

\end{itemize}
In the remainder of this section, we first develop a Bayesian model to support the two key designs mentioned above. We then propose an approximate sparse channel estimation algorithm based on this Bayesian model, using the turbo-VBI technique \cite{turbo-VBI}.

\subsection{Bayesian Modeling}
We start by constructing a Bayesian probability model for the observation $\widetilde{\bf Y}_{\rm u}$ and the estimation variables ${\bm \Psi}_{{\rm V}}$ and ${\bm \Psi}_{{\rm H}}$. Specifically, we propose a likelihood model to facilitate the design of the low-complexity 2D-LMMSE, and a prior model to incorporate the prior mask.

\subsubsection{Likelihood Model}
To approximate the 2D-LMMSE using two 1D-LMMSE estimators, we introduce two auxiliary variable $\overline{\bf X}_{\rm B}={\bf A}_{\rm u} \widetilde{\bm \Psi}$ with ${\bf A}_{\rm u}=\widetilde{\bf U}_{\rm u}^{\rm H}\widetilde{\bf G}_{\rm u}$, and $\overline{\bf X}_{\rm A}=\overline{\bf X}_{\rm B}^{\rm T}$, the observations shown in \eqref{equ:UL_signal_Yu_svd} and \eqref{equ:UL_signal_Yu_svd_2} can be expressed as:
\begin{align}
\widetilde{\bf Y}_{\rm u}
&=\overline{\bf X}_{\rm A}^{\rm T} \overline{\bf F}^{\rm T}+ \widetilde{\bf Z}_{\rm u}, \label{equ:UL_signal_Yu_svd_E} \\
\overline{\bf X}_{\rm A} &= \overline{\bf X}_{\rm B}^{\rm T},\label{equ:UL_signal_E}\\
\overline{\bf X}_{\rm B}&={\bf A}_{\rm u} \widetilde{\bm \Psi}, \label{equ:UL_signal_E_Psi}
\end{align}
where $\widetilde{\bm \Psi}=[{\bm \Psi}_{{\rm V}}^{\rm T},{\bm \Psi}_{{\rm H}}^{\rm T}]^{\rm T}$.
This shows that the 2-D sparse domain transformation has been broken down into two 1-D sparse domain transformations.

According to the new observation model shown in \eqref{equ:UL_signal_Yu_svd_E}, \eqref{equ:UL_signal_E}, and \eqref{equ:UL_signal_E_Psi}, we construct the likelihood model as follows:
\begin{itemize}
\item {\bf Relationship between} $\widetilde{\bf Y}_{\rm u}$ {\bf and} $\overline{\bf X}_{\rm A}$: According to \eqref{equ:UL_signal_Yu_svd_E}, the likelihood for the observation $\widetilde{\bf Y}_{\rm u}$ is given by
    \begin{equation}\label{equ:likelihood_Y}
    \begin{aligned}[b]
    p(\widetilde{\bf Y}_{\rm u} | \overline{\bf X}_{\rm A})={\cal{CMN}}(\overline{\bf X}^{\rm T}_{\rm A} \overline{\bf F}^{\rm T},\sigma^2_{\rm z}{\bf I}_{\widetilde{M}}, {\bf I}_{N_{\rm c}}).
    \end{aligned}
    \end{equation}
where ${\cal{CMN}}({\bm \Omega},{\sigma^2_{J}}{\bf I}_J,{\bf I}_K)$ denotes the PDF of the matrix complex Gaussian distribution whose elements are uncorrelated to each other, and the $(j,k)$-th element follows ${\cal{CN}}({ \Omega}_{j,k},\sigma^2_{J})$.


\item {{\bf Equivalence of} $\overline{\bf X}_{\rm A}$ {\bf and} $\overline{\bf X}_{\rm B}$}: We use the Dirac delta distribution to indicate the equation $\overline{\bf X}_{\rm A}=\overline{\bf X}_{\rm B}^{\rm T}$:
    \begin{equation}\label{equ:likelihood_E_eq}
    \begin{aligned}[b]
    &p(\overline{\bf X}_{\rm A} | \overline{\bf X}_{\rm B})
    =\delta(\overline{\bf X}_{\rm A}-\overline{\bf X}_{\rm B}^{\rm T}).
    \end{aligned}
    \end{equation}

\item {{\bf Relationship between} $\overline{\bf X}_{\rm B}$ {\bf and} $\widetilde{\bm \Psi}$}:
    Similarly, we use the Dirac delta distribution to model the equation $\overline{\bf X}_{\rm B}={\bf A}_{\rm u} \widetilde{\bm \Psi}$:
    \begin{equation}\label{equ:likelihood_E_define}
    \begin{aligned}[b]
    &p(\overline{\bf X}_{\rm B} | \widetilde{\bm \Psi})
    =\delta(\overline{\bf X}_{\rm B}- {\bf A}_{\rm u} \widetilde{\bm \Psi}),
    \end{aligned}
    \end{equation}
    where $\widetilde{\bm \Psi}=[{\bm \Psi}_{{\rm V}}^{\rm T},{\bm \Psi}_{{\rm H}}^{\rm T}]^{\rm T}$.


\end{itemize}

\subsubsection{Prior Model}
We begin by constructing a prior mask $\bf D$, followed by proposing a two-layer prior model for ${\bm \Psi}_{{\rm V}}$ and ${\bm \Psi}_{{\rm H}}$ that incorporates the mask $\bf D$ to aid in magnitude estimation.
\begin{itemize}
\item {\bf Constructing the Prior Mask $\bf D$}: We create the prior mask ${\bf D} \in \{0,1\}^{|{\mathcal B}| \times L}$ based on the support $\cal D$ provided by the modified OMP estimator. The $(b,\ell)$-th element in ${\bf D}$ is defined as follows:
    \begin{equation}\label{eq::mask}
    \begin{aligned}[b]
    d_{\ell,b}=
    \begin{cases}
    1,   &{\rm{if}} \;\;  (b,\ell) \in {\cal D} \; {\rm{or}} \; {\cal D} \cap \overline{\cal D}^{(b,\ell)} \neq \emptyset, \\
    0,  &{\rm{Otherwise}},
    \end{cases}
    \end{aligned}
    \end{equation}
    where $\overline{\cal D}^{(b,\ell)}$ represents the set of all adjacent points for index $(b,\ell)$ in the angular-delay domain. This design leverages the clustering feature of the channel support; specifically, an element in ${\bm \Psi}_{{\rm V}}$ (or ${\bm \Psi}_{{\rm H}}$) is likely to be non-zero if one of its adjacent elements is non-zero.

\item {\bf Layer 1: Bernoulli-Gaussian Prior}: Based on the propagation channel model in \eqref{equ:signal_model_app}, we introduce a Bernoulli-Gaussian prior for the elements in ${\bm \Psi}_{{\rm V}}$ and ${\bm \Psi}_{{\rm H}}$ with common parameters:
    \begin{align}
    p( \psi_{{\rm V},\ell,b} | \alpha_{\ell,b}; d_{\ell,b})
    &=d_{\ell,b}{\cal{CN}}(\psi_{{\rm V},\ell,b}; 0, \alpha_{\ell,b}^{-1}) \notag \\
    &\;\;\;+({1-d_{\ell,b}}) \delta(\psi_{{\rm V},\ell,b}),\label{equ:pri_L1_V}\\
    p( \psi_{{\rm H},\ell,b} | \alpha_{\ell,b};d_{\ell,b})
    &=d_{\ell,b}{\cal{CN}}(\psi_{{\rm H},\ell,b}; 0, \alpha_{\ell,b}^{-1}) \notag \\
    &\;\;\;+({1-d_{\ell,b}}) \delta(\psi_{{\rm H},\ell,b}), \label{equ:pri_L1_H}
    \end{align}
    where $\psi_{{\rm V},\ell,b}$ and $\psi_{{\rm H},\ell,b}$ are the $(b,\ell)$-th elements in ${\bm \Psi}_{{\rm V}}$ and ${\bm \Psi}_{{\rm H}}$. Note that if the mask $d_{\ell,b} =0$, then $\psi_{{\rm V},\ell,b}=\psi_{{\rm H},\ell,b}=0$.
\item {\bf Layer 2: Gamma Prior for the Magnitude Estimation}: We use a Gamma prior for the inverse variance $\alpha_{\ell,b}$:
    \begin{align}
    p(\alpha_{\ell,b} )&={\rm Gam}(a_0, c_0),
    \end{align}
    where $a_0$ is the shape parameter  and $c_0$ is the rate parameter. We set  both $a_0$ and $c_0$ to small values close to zero to create an uninformative prior \cite{VBI2008}. Note that if the mask $d_{\ell,b} =0$, the estimation of $\alpha_{\ell,b}$ is skipped.
\end{itemize}

\begin{figure}
[!t]
\centering
\includegraphics[width=.90\columnwidth]{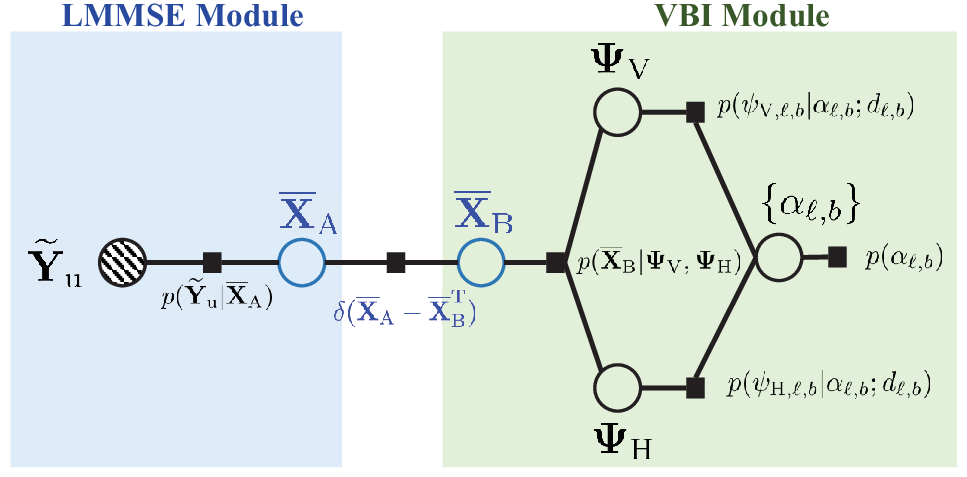}
\caption{Illustration of the constructed Bayesian model.}
\label{factor_graph}
\end{figure}

We summarize the overall Bayesian model as shown in {\figurename~\ref{factor_graph}}. The channel estimation problem involves inferring the posterior mean values of ${\bm \Psi}_{{\rm V}}$ and ${\bm \Psi}_{{\rm H}}$.

\subsection{Turbo-VBI for Approximated Posterior Inference}
We propose a turbo-VBI estimator that performs approximate Bayesian inference on the joint posterior distribution of the variables $\overline{\bf X}_{\rm A}$,  ${\bm \Psi}_{{\rm V}}$, ${\bm \Psi}_{{\rm H}}$, and all $\alpha_{\ell,b}$, given the observation $\widetilde{\bf Y}_{\rm u}$. Specifically, we apply the mean-field assumption, leading to the following joint approximate posterior:
\begin{equation}\label{equ:app_posterior_joint}
\begin{aligned}[b]
q(\overline{\bf X}_{\rm A}, {\bm \xi})&=
q(\overline{\bf X}_{\rm A})q({\bm \Psi}_{{\rm V}})q({\bm \Psi}_{{\rm H}}) \prod_{\ell,b} q(\alpha_{\ell,b}),
\end{aligned}
\end{equation}
where ${\bm \xi}=\{{\bm \Psi}_{{\rm V}}, {\bm \Psi}_{{\rm H}},\{\alpha_{\ell,b}\} \}$ summarizes all the variables involved in the prior model, and $q(\cdot)$ indicates the approximate posterior. As shown in {\figurename~\ref{factor_graph}}, the turbo-VBI solution divides the factor graph into two modules, applying low-complexity Bayesian inference to each module separately. 

\subsubsection{Approximate Message Passing between Two Modules}
By introducing factor node $p(\overline{\bf X}_{\rm A} | \overline{\bf X}_{\rm B})=\delta(\overline{\bf X}_{\rm A}-\overline{\bf X}_{\rm B}^{\rm T})$, the overall joint Bayesian inference is divided into the LMMSE module and the VBI module. The approximate messages exchanged between the two modules are defined by
\begin{itemize}
\item \textbf{Approximate Prior}: This message is sent from the VBI module to the LMMSE module, which is designed as
\begin{equation}\label{equ:app_pri_fix}
\begin{aligned}[b]
\varsigma_{\rm{pri}}(\overline{\bf X}_{\rm A})
={\cal{CMN}}({\bm 0},\overline{\sigma}^2_{\rm{xapri}}{\bf I}_L, {\bf I}_{\widetilde{M}}),
\end{aligned}
\end{equation}
where the variance $\overline{\sigma}^2_{\rm{xapri}}$ is obtained by combining the equation $\overline{\bf X}_{\rm A}=\widetilde{\bm \Psi}^{\rm T} {\bf A}_{\rm u}^{\rm T} $ and the prior of $\widetilde{\bm \Psi}$ given in \eqref{equ:pri_L1_V} and \eqref{equ:pri_L1_H}:
\begin{equation}\label{equ:app_pri_fix_var}
\begin{aligned}[b]
\overline{\sigma}^2_{\rm{xapri}}&=\frac{1}{{\widetilde{M}} {L}} \sum_{\ell=1}^L
{\rm{Tr}}\left( {\bf A}_{\rm u}
{\rm{diag}} \left( \left[ {\bm \kappa}_{\ell}^{\rm T}, {\bm \kappa}_{\ell}^{\rm T} \right]^{\rm T}\right)
 {\bf A}_{\rm u}^{\rm H} \right),
\end{aligned}
\end{equation}
where ${\bm \kappa}_{\ell}=[\kappa_{\ell,1}, \kappa_{\ell,2},\cdots,\kappa_{\ell,|{\mathcal B}|}]^{\rm T}$ with $b$-th element $\kappa_{\ell,b}$ defined by
\begin{equation}\label{equ:app_pri_fix_var_element}
\begin{aligned}[b]
\kappa_{\ell,b}&=  d_{\ell,b} {\mathbb E}_{q(\alpha_{\ell,b})} [\alpha_{\ell,b}^{-1} ]
.
\end{aligned}
\end{equation}

\item \textbf{Approximate Likelihood}: This message is sent from the LMMSE module to the VBI module. The posterior of $\overline{\bf X}_{\rm A}$ is given by
    \begin{align}
    {q}(\overline{\bf X}_{{\rm A}})&=\prod_{m=1}^{\widetilde{M}} {q}(\overline{\bf x}_{{\rm A},m}), \notag \\
    {q}(\overline{\bf x}_{{\rm A},m})
    &={\cal{CN}}({\bm \mu}_{{\rm{xapos}},m},{\bf C}_{{\rm{xapos}},m} ), \label{equ:post_Xa}
    \end{align}
    for $m=1,2,\cdots,\widetilde{M}$. Following the message approximation approach from \cite{VAMP}, we approximate ${q}(\overline{\bf X}_{{\rm A}})$ as
    \begin{equation}\label{equ:post_Xa_app}
    \begin{aligned}[b]
    \overline{q}(\overline{\bf X}_{{\rm A}})
    &={\cal{CMN}}({\bm \Omega}_{{\rm{xapos}}},\overline\sigma_{\rm{xapost}}^2 {\bf I}_L, {\bf I}_{\widetilde{M}} ),
    \end{aligned}
    \end{equation}
    where the parameters are defined as:
    \begin{align}
    {\bm \Omega}_{{\rm{xapos}}}&=[{\bm \mu}_{{\rm{xapos}},1},\cdots,{\bm \mu}_{{\rm{xapos}},\widetilde{M}}],
    \\
    \overline\sigma_{\rm{xapost}}^2
    &=\frac{1}{{\widetilde{M}} {L}} \sum_{m=1}^{\widetilde{M}} {\rm{Tr}}\left({\bf C}_{{\rm{xapos}},m}\right). \label{equ:post_Xa_app_var}
    \end{align}
    The approximate likelihood is obtained by substituting \eqref{equ:app_pri_fix} and \eqref{equ:post_Xa_app} into $\varsigma_{\rm{lik}}=\frac{\overline{q}(\overline{\bf X}_{{\rm A}})}{\varsigma_{\rm{pri}}}$, resulting in:
    \begin{align}
    \varsigma_{\rm{lik}}(\overline{\bf X}_{{\rm B}})&={\cal{CMN}}({\bm \Omega}_{{\rm{xblik}}},\overline\sigma_{\rm{xblik}}^2  {\bf I}_{\widetilde{M}} , {\bf I}_L),  \label{equ:alike_Xa}
    \end{align}
    where the parameters are calculated by:
    \begin{align}
    \overline\sigma_{\rm{xblik}}^2&= \frac{\overline{\sigma}^2_{\rm{xapri}}\overline\sigma_{\rm{xapost}}^2 }{\overline{\sigma}^2_{\rm{xapri}}- \overline\sigma_{\rm{xapost}}^2}, \\
    {\bm \Omega}_{{\rm{xblik}}} &=\frac{\overline{\sigma}^2_{\rm{xapri}} }{\overline{\sigma}^2_{\rm{xapri}}- \overline\sigma_{\rm{xapost}}^2} {\bm \Omega}_{{\rm{xapos}}}^{\rm T} .
    \end{align}

\end{itemize}

\subsubsection{LMMSE Module for the Frequency-Delay Sparse Domain Transformation}
In this module, we focus on the 1-D sparse domain transformation from the frequency to the delay domain based on the joint distribution:
\begin{equation}\label{equ:app_joint_A}
\begin{aligned}[b]
p(\overline{\bf X}_{\rm A})&=
p(\widetilde{\bf Y}_{\rm u} | \overline{\bf X}_{\rm A}) \varsigma_{\rm{pri}}(\overline{\bf X}_{\rm A}),
\end{aligned}
\end{equation}
where the likelihood $p(\widetilde{\bf Y}_{\rm u} | \overline{\bf X}_{\rm A})$ is defined in \eqref{equ:likelihood_Y}, and the prior $\varsigma_{\rm{pri}}(\overline{\bf X}_{\rm A})$  is given in \eqref{equ:app_pri_fix}. Both are complex Gaussian distributions.

Consequently, the posterior of  $\overline{\bf x}_{{\rm A},\ell}$ also follows a complex Gaussian distribution. Denoting the posterior distribution by ${q}(\overline{\bf x}_{{\rm A},m})
    ={\cal{CN}}({\bm \mu}_{{\rm{xapos}},m},{\bf C}_{{\rm{xapos}},m} )$ and letting ${\bm \Omega}_{{\rm{xapos}}}=[{\bm \mu}_{{\rm{xapos}},1},\cdots,{\bm \mu}_{{\rm{xapos}},\widetilde{M}}]$, the unknown parameters can be calculated by the LMMSE estimator:
\begin{align}
{\bf C}_{{\rm{xpos}},\ell}&=  \left(\overline{\sigma}^{-2}_{\rm{xapri}} {\bf I}_{\widetilde{L}}+\overline{\sigma}^{-2}_{\rm z}\overline{\bf F}^{\rm H} \overline{\bf F}\right)^{-1},\label{equ:apost_lik_var}\\
   {\bm \Omega}_{{\rm{xapos}}}
    &=\overline{\sigma}^{-2}_{\rm z} {\bf C}_{{\rm{xpos}},\ell}  \overline{\bf F}^{\rm H}
     \widetilde{\bf Y}_{\rm u}^{\rm T} .\label{equ:apost_lik_mean}
    \end{align}

\subsubsection{VBI Module for the  Delay-Angular Domain Sparse Channel Estimation}
In this module, we infer the approximate posterior for all variables in ${\bm \xi}=\{{\bm \Psi}_{{\rm V}}, {\bm \Psi}_{{\rm H}},\{\alpha_{\ell,b}\} \}$.
The joint distribution is represented as:
\begin{equation}\label{equ:post_vbi_module}
\begin{aligned}[b]
p({\bm \xi};{\bf d})
& = \prod_{\ell} \bigg( \varsigma_{\rm{lik}}(  \widetilde{\bm \psi}_{\ell})
\prod_{b} \big( p(\alpha_{\ell,b} )
\\
& \quad p( \psi_{{\rm V},\ell,b} | \alpha_{\ell,b}; d_{\ell,b}) p( \psi_{{\rm H},\ell,b} | \alpha_{\ell,b}; d_{\ell,b}) \big) \bigg),
\end{aligned}
\end{equation}
where $\varsigma_{\rm{lik}}(  \widetilde{\bm \psi}_{\ell})$ is obtained by combining the approximate likelihood $\varsigma_{\rm{lik}}(\overline{\bf x}_{{\rm B},\ell})$ with the equation $\overline{\bf X}_{\rm B}={\bf A}_{\rm u} \widetilde{\bm \Psi}$:
\begin{equation}\label{equ:lik_vbi_module}
\begin{aligned}[b]
\varsigma_{\rm{lik}}(  \widetilde{\bm \psi}_{\ell})
& = {\mathbb E}_{\overline{\bf x}_{{\rm B},\ell}} \left[\varsigma_{\rm{lik}}(\overline{\bf x}_{{\rm B},\ell})p(\overline{\bf x}_{{\rm B},\ell} | \widetilde{\bm \psi}_\ell) \right]\\
&={\cal{CN}}({\bm \mu}_{{\rm{xblik}},\ell}- {\bf A}_{\rm u} \widetilde{\bm \psi}_{\ell} ,\overline\sigma_{\rm{xblik}}^2 {\bf I}_{\widetilde{M}} ),
\end{aligned}
\end{equation}
where ${\bm \Omega}_{{\rm{xblik}}}=[{\bm \mu}_{{\rm{xblik}},1},\cdots,{\bm \mu}_{{\rm{xblik}},L}]$.

Since all the distributions involved in the true posterior belong to the exponential family, we design the approximate posterior distributions as the conjugate  distributions to simplified the inference:
\begin{align}
{ q}({\bm \psi}_{{\rm V},\ell})
&={\cal{CN}}({\bm \mu}_{{\rm{posV}},\ell},{\rm {diag}} ({\bm \sigma}^2_{{\rm{posV}},\ell} ) ),\label{equ:post_vbi_psiV}\\
{q}({\bm \psi}_{{\rm H},\ell})
&={\cal{CN}}({\bm \mu}_{{\rm{posH}},\ell},{\rm {diag}} ({\bm \sigma}^2_{{\rm{posH}},\ell} ) ),\label{equ:post_vbi_psiH}\\
q (\alpha_{\ell,b})&={\rm Gam}(\widetilde{a}_{\ell,b}, \widetilde{c}_{\ell,b}), \label{equ:post_vbi_gamma_bell}
\end{align}
for all $b=1,2,\cdots,|{\cal B}|$ and $\ell=1,2,\cdots,L$.

The target approximate posterior inference is usually achieved by minimizing the KL divergence between  $q({\bm \xi})$ and $p({\bm \xi};{\bf d})$, which is equivalent to maximizing the Evidence Lower Bound (ELBO) \cite{VBI2008}. 
Letting ${\bm \varpi}= \{ {\bm \mu}_{{\rm{posV}},\ell},{\bm \sigma}^2_{{\rm{posV}},\ell},{\bm \mu}_{{\rm{posH}},\ell},{\bm \sigma}^2_{{\rm{posH}},\ell},  \widetilde{a}_{\ell,b}, \widetilde{c}_{\ell,b}, \forall b,\ell\} $ summarize all the parameters to be learnt, the ELBO maximization problem is given by
\begin{subequations}
\begin{align}
{\mathcal{P}}{(\text{C})} \quad\; {\bm \varpi}^\ast
& = \underset{ {\bm \varpi} }{\mathrm{ argmax}} \:\:
f_{\rm{ELO}}( {\bm \varpi} ) \notag
\end{align}
\end{subequations}
where
the ELBO function is given by
\begin{equation}\label{eq::elbo}
\begin{aligned}[b]
&f_{\rm{ELO}}( {\bm \varpi} )=
{\mathbb E}_{ q ({\bm \xi}; {\bm \varpi}) } \log \left[ \frac{p({\bm \xi};{\bf d})}
{q({\bm \xi};{\bm \varpi})} \right].
\end{aligned}
\end{equation}

We solve ${\mathcal{P}}{(\text{C})}$ by decoupling the variables into blocks, updating each block separately while fixing the variables in the other blocks.  The detailed updating steps are elaborated as follows:
\begin{itemize}
\item {\bf Dimension Reduced LMMSE}: In this block, we estimate the posteriors of ${\bm \psi}_{{\rm V},\ell}$ and ${\bm \psi}_{{\rm H},\ell}$ for all $\ell=1,\cdots,L$ given $q (\alpha_{\ell,b})$, which can be achieved by the LMMSE estimator. However, the complexity of the LMMSE estimator can be high due to the large dimension of the sensing matrix ${\bf A}_{\rm u} \in {\mathbb C}^{\widetilde{M} \times 2|{\mathcal B}|}$. Fortunately, if the mask $d_{\ell,b} =0$, then $\psi_{{\rm V},\ell,b}=\psi_{{\rm H},\ell,b}=0$. Thus, we can simplify the process by skipping the estimation of all zero-masked variables. We define the set ${\cal B}_\ell=\{ b | d_{\ell,b}=1, \forall b=1,\cdots,|{\cal B}|\}$. Then, ${\bf A}_{{\rm u},\ell} $ selects the active $2|{\cal B}_\ell|$ columns of ${\bf A}_{\rm u}$ based on ${\cal B}_\ell$, and the dimension-reduced LMMSE is given by:
    \begin{align}
{\bf C}_{{\rm{eff}},\ell}&=  \left( {\rm{diag}}([1,1]^{\rm T} \otimes {\overline{\bm \kappa}_\ell})^{-1}+\overline\sigma_{\rm{xblik}}^{-2} {\bf A}_{{\rm u},\ell}^{\rm H} {\bf A}_{{\rm u},\ell} \right)^{-1},\label{equ:post_vbi_psi_var}\\
   {\bm \mu}_{{\rm{eff}},\ell}
    &=\overline\sigma_{\rm{xblik}}^{-2}  {\bf C}_{{\rm{eff}},\ell}  {\bf A}_{{\rm u},\ell}^{\rm H}
     {\bm \mu}_{{\rm{xblik}},\ell}, \label{equ:post_vbi_psi_mean}
    \end{align}
    where $\overline{\bm \kappa}_\ell$ selects the $|{\cal B}_\ell|$ elements from ${\bm \kappa}_\ell$ based on ${\cal B}_\ell$. Finally, we map ${\bm \mu}_{{\rm{eff}},\ell}$ and ${\rm{diag}}({\bf C}_{{\rm{eff}},\ell})$ into ${\bm \mu}_{{\rm{posV}},\ell}$, ${\bm \sigma}^2_{{\rm{posV}},\ell}$, ${\bm \mu}_{{\rm{posH}},\ell}$, and ${\bm \sigma}^2_{{\rm{posH}},\ell}$ according to the index set ${\cal B}_\ell$, while setting the remaining elements to zero.

\item {\bf Inverse Variance Estimation}: In this module, we estimate $q (\alpha_{\ell,b})$ to account for the magnitude variation of the non-zero variables. Specifically, we update the parameters in $q (\alpha_{\ell,b})$ if $d_{\ell,b}=1$:
    \begin{align}
    \widetilde{a}_{\ell,b}&= a_0+2  , \label{equ:post_vbi_a_bl} \\
    \widetilde{c}_{\ell,b}&= c_0+ {|\mu|}^2_{{\rm{posH}},\ell,b}+{|\mu|}^2_{{\rm{posV}},\ell,b} \notag\\
    &\quad+{\sigma}^2_{{\rm{posV}},\ell,b}+{\sigma}^2_{{\rm{posH}},\ell,b}. \label{equ:post_vbi_c_bl}
    \end{align}
    The expectation value of $\alpha_{\ell,b}$ is ${\mathbb E}[\alpha_{\ell,b}]=\frac{\widetilde{a}_{\ell,b}}{\widetilde{c}_{\ell,b}}$.
\end{itemize}


\subsubsection{Summary of Algorithm}

The proposed turbo-VBI estimation algorithm is summarized in algorithm \ref{alg:Turbo-VBI}.
The algorithm is initialized by the modified OMP to obtain the mask $D$, and the additional complexity is about
${\cal O}\left(  \widetilde{M}N_{\rm c}L^2+\widetilde{M} L^3+\sum_{\ell} \widetilde{M} |{\cal B}_\ell|^2 +\sum_\ell |{\cal B}_\ell|^3\right)$. 


\begin{algorithm}[!t]
\caption{Proposed Turbo-VBI Estimator}\label{alg:Turbo-VBI}
\begin{algorithmic}[1]
\State\textbf{Input}:  Observation  $\widetilde{\bf Y}_{\rm u}$ and mask $\bf D$;
\State \textbf{Initialize}: ${\bm \varpi}$, $\varsigma_{\rm{pri}}(\overline{\bf X}_{\rm A})$, and $I_{\rm{max}}$;
\For{$i = 1$ {\textbf{to}} $I_{\rm{max}}$}
    \State \%\% {\emph{LMMSE Module}}:
    \For{$m = 1$ {\textbf{to}} $\widetilde{M}$}
        \State Update ${\bf C}_{{\rm{xpos}},\ell}$ and ${\bm \Omega}_{{\rm{xapos}}}$ using \eqref{equ:apost_lik_var} and \eqref{equ:apost_lik_mean}.
    \EndFor

    \State Calculate approximate likelihoods $\varsigma_{\rm{lik}}(\overline{\bf X}_{{\rm B}})$ in \eqref{equ:alike_Xa};
    \State \%\% {\emph{VBI Module}}:
    \For{$\ell = 1$ {\textbf{to}} $L$}
        \State Update ${ q}({\bm \psi}_{{\rm V},\ell})$ and ${ q}({\bm \psi}_{{\rm H},\ell})$ by \eqref{equ:post_vbi_psi_var} and \eqref{equ:post_vbi_psi_mean};
        \State update $q (\alpha_{\ell,b})$ by \eqref{equ:post_vbi_a_bl} and \eqref{equ:post_vbi_c_bl};
    \EndFor
    \State Update approximate prior $\varsigma_{\rm{pri}}(\overline{\bf X}_{\rm A})$ by \eqref{equ:app_pri_fix_var}.
\EndFor
\State\textbf{Output}: The estimated ${\bm \Psi}_{\rm V}$ and ${\bm \Psi}_{\rm H}$.
\end{algorithmic}
\end{algorithm}

\section{Analog Precoder Optimization}
In this section, we solve ${\mathcal{P}}{(\text{A})}$ to optimize analog precoder (i.e., the FAS state vector $\bf s$)  by solving ${\mathcal{P}}{(\text{A})}$ given estimated ${\bm \Psi}_{{\rm V},k}$ and ${\bm \Psi}_{{\rm H},k}$ for all $K$ users. Recall that the problem formulation is given by
\begin{align}
    {\mathcal{P}}{(\text{A})}
  \;\;\; {\bf{s}}^\ast  & =\; {\rm{argmax}}_{  \bf{s}
 } \; \;
\frac{1}{N_{\rm c}}\sum_{n_{\rm c}=1}^{N_{\rm c}} {R}_{n_{\rm c},k}({\bf s}) \notag\\
 & \; {\text{s.t.}} \;  s_m \in \{1,2,\cdots,N_{\rm s}\}, \; \forall  m, \label{equ:discrete_1_2}
\end{align}
where ${R}_{n_{\rm c},k}({\bf s})=\log_2 \left( 1+ \frac{\gamma_{n_{\rm c}}({\bf s})}{\sigma_{\overline{\rm z}}^2} \right)$, and the SNR ${\gamma_{n_{\rm c}}({\bf s}^\ast)}$ is defined by
\begin{equation}\label{equ:power_zf_2}
\begin{aligned}[b]
{\gamma_{n_{\rm c}}({\bf s}^\ast)}= \frac{  K  P_{\rm T}} {  {\rm Tr} \left(
\left({\bf H}^{\rm T}_{n_{\rm c}}({\bf s}^\ast) {\bf H}^\star_{n_{\rm c}}({\bf s}^\ast) \right)^{-1}
\right)}.
\end{aligned}
\end{equation}

\subsection{Dimension Reduced Channel Model}
In contrast to using the channel model in \eqref{equ:signal_model_app} with all rays in the angular grid set, we adopt a simplified model that incorporates the estimated channel support $\cal D$ (or mask $\bf D$) to reduce computational complexity:
\begin{equation}\label{equ:signal_model_app_2}
\begin{aligned}[b]
&{\bf h}_{n_{\rm c},k}({\bf s}) =
\sum_{ \{b,\ell\} \in {\cal D}}
\bigg( {\rm{diag}} \big(
 {\bm \nu}_{\rm V}(
 {\bm \phi}^{(b)} ; {\bf s}
 )  \big) {\bm \beta}(
 {\bm \phi}^{(b)})\psi_{{\rm V},k}^{(b,\ell)} \\
 & \;\;\;
  + {\rm{diag}} \big(
 {\bm \nu}_{\rm H} (
 {\bm \phi}^{(b)} ; {\bf s}
 )  \big) {\bm \beta}(
 {\bm \phi}^{(b)})
\psi_{{\rm H},k}^{(b,\ell)}  \bigg)
{e^{ - \jmath \frac{{2\pi {\tau_{\ell}} \left( {n_{\rm c} - 1} \right)}}{{{N_{\rm{c}}}}}}}.
\end{aligned}
\end{equation}

\subsection{Reformulated ${\mathcal{P}}{(\overline{\text{A}})}$ for Low-complexity Solution}
The problem ${\mathcal{P}}{(\text{A})}$ is hard to solve because of the discrete constraint $s_m \in \{1,2,\cdots,N_{\rm s}\}$ in \eqref{equ:discrete_1_2}. To address this issue, we suggest a three-step approximation to make the constraint easier to handle.
\begin{itemize}
\item {\bf Reparameterization by Using One-Hot Vectors}: We define a one-hot vector $\overline{\bf s}_m=[\overline{s}_{m,1},\overline{s}_{m,2},\cdots,\overline{s}_{m,N_{\rm s}}]^{\rm T}$ for the P-FAS state configuration. Here $\overline{s}_{m,n_{\rm s}} \in \{0,1\}$ and $\sum_{n_{\rm s}} \overline{s}_{m,n_{\rm s}}=1$.
    We can define new functions for the radiation patterns of the $m$-th antenna as:
    \begin{align}
        \overline{\nu}_{\rm V}({\bm \phi}^{(b)} ; \overline{s}_m)&=\sum_{i=1}^{N_{\rm s}}
\overline{s}_{m,n} {\nu}_{\rm V}({\bm \phi}^{(b)} ; i),\label{equ:resp_one_hot_V}\\
    \overline{\nu}_{\rm H}({\bm \phi}^{(b)} ; \overline{s}_m)&=\sum_{i=1}^{N_{\rm s}}
    \overline{s}_{m,n} {\nu}_{\rm H}({\bm \phi}^{(b)} ; i), \label{equ:resp_one_hot_H}
    \end{align}
    and obviously we have
    \begin{align}
        {\nu}_{\rm V}({\bm \phi}^{(b)} ; {s}_m)&=\overline{\nu}_{\rm V}({\bm \phi}^{(b)} ; \overline{s}_m),\label{equ:resp_one_hot_eq_V}\\
    {\nu}_{\rm H}({\bm \phi}^{(b)} ; {s}_m)&=\overline{\nu}_{\rm H}({\bm \phi}^{(b)} ; \overline{s}_m). \label{equ:resp_one_hot_eq_H}
    \end{align}
    Both equations \eqref{equ:resp_one_hot_V} and \eqref{equ:resp_one_hot_H} do not depend on the instantaneous channel responses ${\bm \Psi}_{{\rm V},k}$ and ${\bm \Psi}_{{\rm H},k}$. Therefore, they can be prepared by offline computation.

\item {\bf Continuous Relaxation with Sparse Promoting Constraints}: The one-hot vector turns the optimization problem into zero-one integer programming, but the constraints are still discrete. To solve this, we relax the constraints on $\overline{s}_{m,n_{\rm s}}$ to a continuous set:
    \begin{align}
    &\sum_{n_{\rm s}} \sqrt{ \overline{s}_{m,n_{\rm s}} }=1, \label{equ:continuous_1} \\
      &  0 \leq \overline{s}_{m,n_{\rm s}} \leq 1. \label{equ:continuous_2}
    \end{align}
    The main idea is to prefer larger values of $\overline{s}_{m,n_{\rm s}}$  to increase the amplitude of the channel impulse response. However, $\overline{s}_{m,n_{\rm s}}$ cannot exceed  $\sqrt{ \overline{s}_{m,n_{\rm s}} }$, with equality only when $\overline{s}_{n,\ell} \in \{0,1\}$. Thus, constraint \eqref{equ:continuous_1} promotes sparsity in $\overline{s}_{m,n_{\rm s}}$, encouraging values close to $0$ or $1$.

\item {\bf Transformation to Unconstrained Problem with Latent Variables}:
    In practice, it's easier to work with an unconstrained problem, which can be solved using tools like PyTorch \cite{NEURIPS2019_9015} and TensorFlow \cite{abadi2016tensorflow}. To achieve this, we introduce a latent variable $\widetilde{\bf s}_m$, related to $\overline{\bf s}_m$ by
    \begin{equation}\label{equ:s_constraint_repar}
    \begin{aligned}[b]
    \overline{\bf s}_m=\left({\rm Softmax}(\widetilde{\bf s}_m)\right)^2.
    \end{aligned}
    \end{equation}
    This relationship incorporates the constraints from \eqref{equ:continuous_1} and \eqref{equ:continuous_2} implicitly.

\end{itemize}

Finally, we define $\widetilde{\bf S}=[\widetilde{\bf s}_1,\widetilde{\bf s}_2,\cdots,\widetilde{\bf s}_M]$ and approximate ${\mathcal{P}}{(\text{A})}$ by following tractable problem:
\begin{align}
    {\mathcal{P}}{(\overline{\text{A}})}
  \;\;\; \widetilde{\bf S}^\ast  & =\; {\rm{argmax}}_{  \widetilde{\bf S}
 } \; \;
 \sum_{n_{\rm c}} {R}_{n_{\rm c},k}(\widetilde{\bf S}) \notag\\
 & \; {\text{s.t.}} \;  \eqref{equ:downlink_rate_2}, \eqref{equ:power_zf_2}, \eqref{equ:signal_model_app_2},
 \eqref{equ:resp_one_hot_V}, \eqref{equ:resp_one_hot_H}, \eqref{equ:resp_one_hot_eq_V}, \eqref{equ:resp_one_hot_eq_H}, \eqref{equ:s_constraint_repar}.
 \notag
\end{align}
${\mathcal{P}}{(\overline{\text{A}})}$ is an unconstrained problem because all the expressions in the ``subject to'' block are equations. We solve ${\mathcal{P}}{(\overline{\text{A}})}$ using the Adam optimizer \cite{Adam} in PyTorch with a learning rate of $0.001$.

\begin{figure}
[!t]
\centering
\includegraphics[width=0.99\columnwidth]{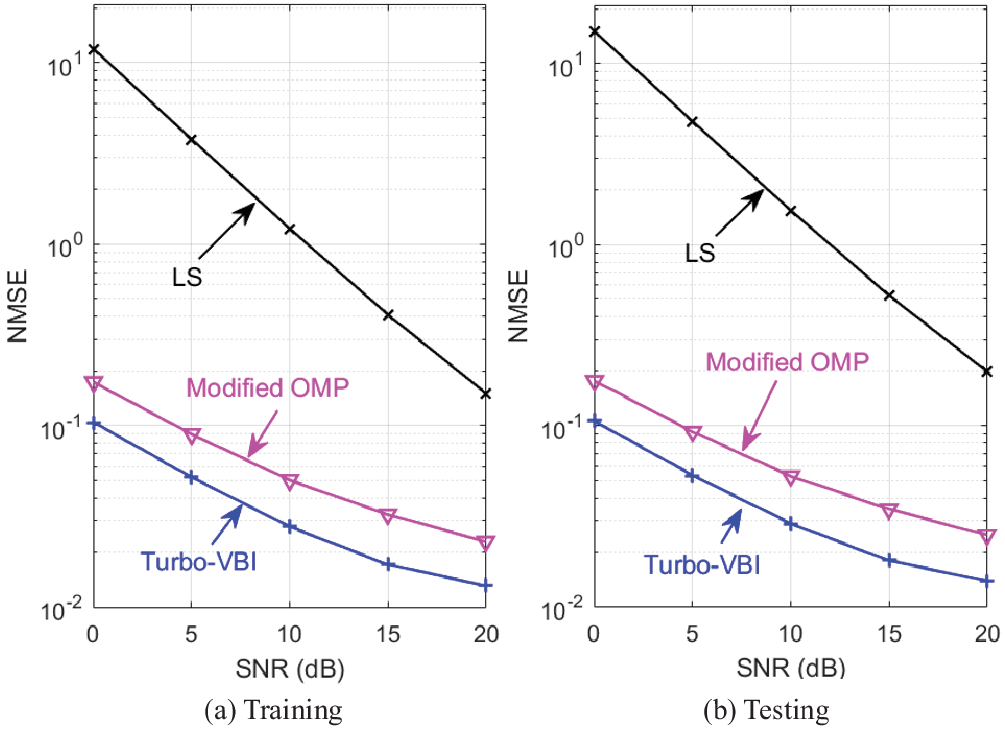}
\caption{NMSE versus SNR for different channel estimation algorithms.}
\label{CE_eva}
\end{figure}

\section{Simulations}
We examine an indoor femtocell network at 2.5 GHz, using the COST2100 model for channel realizations \cite{cost2100@twc}. The BS is at the origin and employs $M=16$ P-FAS antennas \cite{Pixelantenna2025}. Users are randomly located in a circle centered at (10 m, 0 m) with a radius of 5 m. The OFDM length is $N_{\rm c}=256$ with a subcarrier spacing of 15 kHz. Each user has $T=4$ uplink channel sounding blocks with random FAS states, and each OFDM pilot is shared among 4 users using the comb-4 SRS scheme in 5G NR. The overall uplink pilot overhead is $4 \lfloor \frac{K}{4} \rfloor$ OFDM symbols for $K$ users. For simplicity, we set the noise power to $\sigma_{\rm z}^2=\sigma_{\overline{\rm z}}^2=1$, and each user's uplink transmit power for each subcarrier is  $P_{\rm T}$ as well. To evaluate the effectiveness of the proposed solution, we consider the following baselines:
\begin{itemize}
\item {\bf Random}: Each antenna selects a random state. This baseline does not require channel estimation for FAS configuration.
\item {\bf Non-FAS}: This scheme employs non-fluid antennas with equal magnitudes of antenna responses for $0\leq \theta \leq \pi/2$ and $0\leq \varphi < 2\pi$ with normalized average antenna gain.
\item {\bf Group-Opt \cite{greedyoptStatisticCE2023WSA,greedyopt2019TWC}}: In this scheme, all the antennas select the same state, and the optimal state is selected by exhaustive search given perfect CSI.
\item {\bf Upper Bound}: This scheme solves ${\mathcal{P}}{(\overline{\text{A}})}$ under the assumption that the effective channels for any FAS state configurations are perfectly known.
\end{itemize}

{\figurename~\ref{CE_eva}} illustrates the estimation NMSE versus SNR for various channel estimation algorithms. We evaluate the accuracy of channels resulting from FAS state configurations during uplink channel sounding, as well as the accuracy of the predicted channel using 50 random FAS configurations. The results show that the proposed grid-based approximate separable channel model enables consistent performance for both training and testing.
Additionally, the LS algorithm performs poorly because it ignores channel sparsity, while the modified OMP algorithm achieves good performance at high SNRs. The proposed Turbo-VBI algorithm further enhances performance by about 5 dB, thanks to its variance estimation for the non-zero elements.

\begin{figure}
[!t]
\centering
\includegraphics[width=0.99\columnwidth]{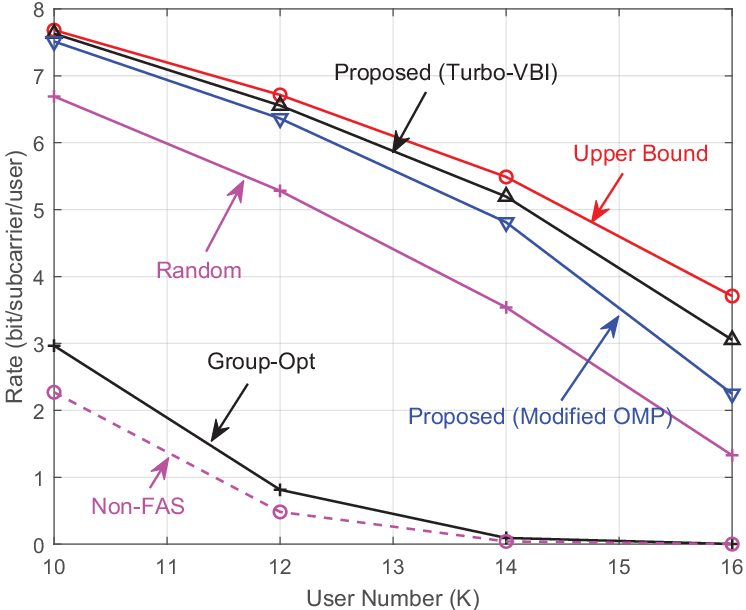}
\caption{Achievable rate versus number of users for different channel schemes with $P_{\rm T}=20$ dB.}
\label{rate_vs_K}
\end{figure}

{\figurename~\ref{rate_vs_K}} shows the achievable rate versus the number of users for different schemes with a fixed $P_{\rm T}=20$ dB. The Group-Opt scheme performs slightly better than the Non-FAS baseline, as it selects the optimal antenna pattern based on instantaneous CSI. Additionally, using different states for different antennas leads to significant gains, which is the main motivation for developing the FAS. Our proposed schemes also outperform the random state baseline by more than 1 bit/subcarrier/user. Moreover, improving channel estimation accuracy is crucial when the number of users $K$ approaches the number of antennas $M$ to ensure robust precoding gains.

\begin{figure}
[!t]
\centering
\includegraphics[width=0.99\columnwidth]{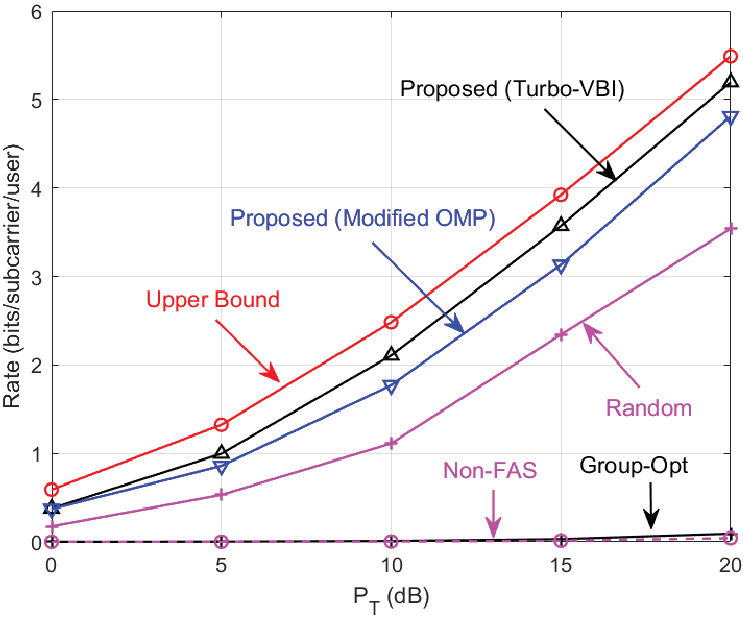}
\caption{Achievable rate versus $P_{\rm T}$ for different schemes with $K=14$.}
\label{rate_vs_snr}
\end{figure}

{\figurename~\ref{rate_vs_snr}} shows the achievable rate versus $P_{\rm T}$ for different schemes with $K=14$. Both the Group-Opt and Non-FAS baselines perform poorly due to unfavorable channel conditions and a high number of users. In contrast, our proposed scheme with Turbo-VBI channel estimation achieves over 5 dB gains compared to the random state baseline, with performance improving as $P_{\rm T}$ increases.

\section{Conclusion}\label{conclusion}
This paper presents an uplink channel estimation and downlink analog precoding scheme for a multiuser MIMO-OFDM system using pixel-based fluid antennas, based on measurements from a real-world prototype. We first introduce a DNN-based approach to approximate the complex antenna radiation pattern functions for various switching states. Building on this, we develop an approximate separable channel response model and design two sparse channel recovery algorithms that achieve high prediction accuracy across all fluid antenna state configurations. Finally, we propose a low-complexity analog precoding algorithm to optimize the switching states for different antennas. Simulation results demonstrate that our solution significantly outperforms various baseline schemes, particularly in high SNR scenarios with a large number of users.

\bibliographystyle{IEEEtran}
\bibliography{IEEEabrv, bibliography}

\begin{thebibliography}{10}
\providecommand{\url}[1]{#1}
\csname url@samestyle\endcsname
\providecommand{\newblock}{\relax}
\providecommand{\bibinfo}[2]{#2}
\providecommand{\BIBentrySTDinterwordspacing}{\spaceskip=0pt\relax}
\providecommand{\BIBentryALTinterwordstretchfactor}{4}
\providecommand{\BIBentryALTinterwordspacing}{\spaceskip=\fontdimen2\font plus
\BIBentryALTinterwordstretchfactor\fontdimen3\font minus
  \fontdimen4\font\relax}
\providecommand{\BIBforeignlanguage}[2]{{%
\expandafter\ifx\csname l@#1\endcsname\relax
\typeout{** WARNING: IEEEtran.bst: No hyphenation pattern has been}%
\typeout{** loaded for the language `#1'. Using the pattern for}%
\typeout{** the default language instead.}%
\else
\language=\csname l@#1\endcsname
\fi
#2}}
\providecommand{\BIBdecl}{\relax}
\BIBdecl

\bibitem{FAS2020TWC@KKWong}
K.-K. Wong, A.~Shojaeifard, K.-F. Tong, and Y.~Zhang, ``Fluid antenna
  systems,'' \emph{IEEE Trans. Wireless Commun.}, vol.~20, no.~3, pp.
  1950--1962, 2020.

\bibitem{FAS2023WCL@KKWong}
K.~K. Wong, W.~K. New, X.~Hao, K.~F. Tong, and C.~B. Chae, ``Fluid antenna
  system—part {I}: Preliminaries,'' \emph{IEEE Commun. Lett.}, vol.~27,
  no.~8, pp. 1919--1923, 2023.

\bibitem{Fluid2025Survey}
W.~K. New, K.~K. Wong, H.~Xu, C.~Wang, F.~R. Ghadi, J.~Zhang, J.~Rao, R.~Murch,
  P.~Ramírez-Espinosa, D.~Morales-Jimenez, C.~B. Chae, and K.~F. Tong, ``A
  tutorial on fluid antenna system for {6G} networks: Encompassing
  communication theory, optimization methods and hardware designs,'' \emph{IEEE
  Commun. Surveys Tuts.}, pp. 1--1, 2024.

\bibitem{movable2025Survey}
L.~Zhu, W.~Ma, W.~Mei, Y.~Zeng, Q.~Wu, B.~Ning, Z.~Xiao, X.~Shao, J.~Zhang, and
  R.~Zhang, ``A tutorial on movable antennas for wireless networks,''
  \emph{IEEE Commun. Surveys Tuts.}, 2025.

\bibitem{MovableOPT2023TWC}
L.~Zhu, W.~Ma, B.~Ning, and R.~Zhang, ``Movable-antenna enhanced multiuser
  communication via antenna position optimization,'' \emph{IEEE Trans. Wireless
  Commun.}, vol.~23, no.~7, pp. 7214--7229, 2023.

\bibitem{MovableOPT2025JSAC}
X.~Shao, R.~Zhang, Q.~Jiang, and R.~Schober, ``{6D} movable antenna enhanced
  wireless network via discrete position and rotation optimization,''
  \emph{IEEE J. Sel. Areas Commun.}, 2025.

\bibitem{sumrateMFAS2024CL}
Z.~Cheng, N.~Li, J.~Zhu, X.~She, C.~Ouyang, and P.~Chen, ``Sum-rate
  maximization for fluid antenna enabled multiuser communications,'' \emph{IEEE
  Commun. Lett.}, 2024.

\bibitem{MUMFASopt2024WCL}
H.~Qin, W.~Chen, Z.~Li, Q.~Wu, N.~Cheng, and F.~Chen, ``Antenna positioning and
  beamforming design for fluid antenna-assisted multi-user downlink
  communications,'' \emph{IEEE Wireless Commun. Lett.}, vol.~13, no.~4, pp.
  1073--1077, 2024.

\bibitem{Pixelantenna2025}
J.~Zhang, J.~Rao, Z.~Li, Z.~Ming, C.~Y. Chiu, K.~K. Wong, K.~F. Tong, and
  R.~Murch, ``A novel pixel-based reconfigurable antenna applied in fluid
  antenna systems with high switching speed,'' \emph{IEEE Open J. Antennas
  Propag.}, vol.~6, no.~1, pp. 212--228, 2025.

\bibitem{movableCE2024WCL}
B.~Xu, Y.~Chen, Q.~Cui, X.~Tao, and K.-K. Wong, ``Sparse bayesian
  learning-based channel estimation for fluid antenna systems,'' \emph{IEEE
  Wireless Commun. Lett.}, 2024.

\bibitem{movableCE2023CL}
W.~Ma, L.~Zhu, and R.~Zhang, ``Compressed sensing based channel estimation for
  movable antenna communications,'' \emph{IEEE Commun. Lett.}, 2023.

\bibitem{movableCEBayesian2024TWC}
Z.~Zhang, J.~Zhu, L.~Dai, and R.~W. Heath, ``Successive {Bayesian}
  reconstructor for channel estimation in fluid antenna systems,'' \emph{IEEE
  Trans. Wireless Commun.}, 2024.

\bibitem{movableCEoversample2025TWC}
W.~K. New, K.~K. Wong, H.~Xu, F.~R. Ghadi, R.~Murch, and C.~B. Chae, ``Channel
  estimation and reconstruction in fluid antenna system: Oversampling is
  essential,'' \emph{IEEE Trans. Wireless Commun.}, vol.~24, no.~1, pp.
  309--322, 2025.

\bibitem{sparseCE2010magazine}
C.~R. Berger, Z.~Wang, J.~Huang, and S.~Zhou, ``Application of compressive
  sensing to sparse channel estimation,'' \emph{IEEE Commun. Mag.}, vol.~48,
  no.~11, pp. 164--174, 2010.

\bibitem{burseLASSO2016TWC}
A.~Liu, V.~K.~N. Lau, and W.~Dai, ``Exploiting burst-sparsity in massive {MIMO}
  with partial channel support information,'' \emph{IEEE IEEE Trans. Commun.},
  vol.~15, no.~11, pp. 7820--7830, 2016.

\bibitem{MarkovpriorTVT2017}
L.~Chen, A.~Liu, and X.~Yuan, ``Structured turbo compressed sensing for massive
  {MIMO} channel estimation using a {Markov} prior,'' \emph{IEEE Trans. Veh.
  Technol.}, vol.~67, no.~5, pp. 4635--4639, 2017.

\bibitem{kernelCE2024TC}
K.~Ying, Z.~Gao, Y.~Su, T.~Qin, M.~Matthaiou, and R.~Schober, ``Reconfigurable
  massive {MIMO}: Precoding design and channel estimation in the
  electromagnetic domain,'' \emph{IEEE Trans. Commun.}, vol.~73, no.~5, pp.
  3423--3440, 2025.

\bibitem{FASCERay2023CL}
H.~Xu, G.~Zhou, K.-K. Wong, W.~K. New, C.~Wang, C.-B. Chae, R.~Murch, S.~Jin,
  and Y.~Zhang, ``Channel estimation for {FAS}-assisted multiuser mmwave
  systems,'' \emph{IEEE Commun. Lett.}, 2023.

\bibitem{greedyoptStatisticCE2023WSA}
F.~Armandoust, E.~Tohidi, M.~Kasparick, L.~Wang, A.~H. Gokceoglu, and
  S.~Stanczak, ``{MIMO} systems with reconfigurable antennas: Joint channel
  estimation and mode selection,'' in \emph{n Proc. WSA SCC 26th Int. ITG
  Workshop Smart Antennas 13th Conf. Syst. Commun., Coding}, Feb. 2023, pp.
  1--6.

\bibitem{optStatistic2023CL}
Y.~Ye, L.~You, J.~Wang, H.~Xu, K.-K. Wong, and X.~Gao, ``Fluid antenna-assisted
  {MIMO} transmission exploiting statistical {CSI},'' \emph{IEEE Commun.
  Lett.}, vol.~28, no.~1, pp. 223--227, 2023.

\bibitem{AMP}
D.~L. Donoho, A.~Maleki, and A.~Montanari, ``Message passing algorithms for
  compressed sensing: {I.} motivation and construction,'' in \emph{Proc. IEEE
  ITW}, 2010, pp. 1--5.

\bibitem{VAMP}
S.~Rangan, P.~Schniter, and A.~K. Fletcher, ``Vector approximate message
  passing,'' \emph{IEEE Trans. Inf. Theory}, vol.~65, no.~10, pp. 6664--6684,
  2019.

\bibitem{OAMP}
J.~Ma and L.~Ping, ``Orthogonal amp,'' \emph{IEEE Access}, vol.~5, pp.
  2020--2033, 2017.

\bibitem{TurboOMP2015SPL}
J.~Ma, X.~Yuan, and L.~Ping, ``On the performance of turbo signal recovery with
  partial {DFT} sensing matrices,'' \emph{IEEE Signal Process. Lett.}, vol.~22,
  no.~10, pp. 1580--1584, 2015.

\bibitem{greedyopt2019TWC}
M.~Hasan, I.~Bahceci, and B.~A. Cetiner, ``Downlink multi-user {MIMO}
  transmission for radiation pattern reconfigurable antenna systems,''
  \emph{IEEE Trans. Wireless Commun.}, vol.~17, no.~10, pp. 6448--6463, 2018.

\bibitem{kernelfitting2023Gcom}
K.~Ying and Z.~Gao, ``Precoding design of reconfigurable massive mimo in the
  electromagnetic domain,'' in \emph{Proc. IEEE GLOBECOM}, 2023, pp.
  5769--5774.

\bibitem{kernel2010TSP}
M.~Costa, A.~Richter, and V.~Koivunen, ``Unified array manifold decomposition
  based on spherical harmonics and {2-D} fourier basis,'' \emph{IEEE Trans.
  Signal Process.}, vol.~58, no.~9, pp. 4634--4645, 2010.

\bibitem{OMP}
T.~T. Cai and L.~Wang, ``Orthogonal matching pursuit for sparse signal recovery
  with noise,'' \emph{IEEE Trans. Inf. Theory}, vol.~57, no.~7, pp. 4680--4688,
  2011.

\bibitem{VBI2008}
D.~G. Tzikas, A.~C. Likas, and N.~P. Galatsanos, ``The variational
  approximation for {Bayesian} inference,'' \emph{IEEE Signal Process. Mag.},
  vol.~25, no.~6, pp. 131--146, 2008.

\bibitem{turbo-VBI}
A.~Liu, G.~Liu, L.~Lian, V.~K.~N. Lau, and M.~Zhao, ``Robust recovery of
  structured sparse signals with uncertain sensing matrix: A turbo-{VBI}
  approach,'' \emph{IEEE Trans. Wireless Commun.}, vol.~19, no.~5, pp.
  3185--3198, 2020.

\bibitem{Adam}
D.~P. Kingma and J.~Ba, ``Adam: A method for stochastic optimization,''
  \emph{arXiv preprint arXiv:1412.6980}, 2014.

\bibitem{liquid1}
C.~Borda-Fortuny, L.~Cai, K.~F. Tong, and K.-K. Wong, ``Low-cost 3d-printed
  coupling-fed frequency agile fluidic monopole antenna system,'' \emph{IEEE
  Access}, vol.~7, pp. 95\,058--95\,064, 2019.

\bibitem{liquid2}
\BIBentryALTinterwordspacing
E.~Motovilova and S.~Y. Huang, ``A review on reconfigurable liquid dielectric
  antennas,'' \emph{Materials}, vol.~13, no.~8, 2020. [Online]. Available:
  \url{https://www.mdpi.com/1996-1944/13/8/1863}
\BIBentrySTDinterwordspacing

\bibitem{surface-wave1}
Y.~Shen, K.-F. Tong, and K.-K. Wong, ``Radiation pattern diversified
  double-fluid-channel surface-wave antenna for mobile communications,'' in
  \emph{IEEE Proc. APWC}, 2022, pp. 085--088.

\bibitem{droplet}
\BIBentryALTinterwordspacing
R.~Malinowski, I.~P. Parkin, and G.~Volpe, ``Advances towards programmable
  droplet transport on solid surfaces and its applications,'' \emph{Chem. Soc.
  Rev.}, vol.~49, pp. 7879--7892, 2020. [Online]. Available:
  \url{http://dx.doi.org/10.1039/D0CS00268B}
\BIBentrySTDinterwordspacing

\bibitem{s-FAMA}
K.-K. Wong, D.~Morales-Jimenez, K.-F. Tong, and C.-B. Chae, ``Slow fluid
  antenna multiple access,'' \emph{IEEE Trans. Commun.}, vol.~71, no.~5, pp.
  2831--2846, 2023.

\bibitem{BruceLee}
\BIBentryALTinterwordspacing
K.-K. Wong, K.-F. Tong, Y.~Shen, Y.~Chen, and Y.~Zhang, ``Bruce lee-inspired
  fluid antenna system: Six research topics and the potentials for 6g,''
  \emph{Frontiers Commun. Netw.}, vol.~3, 2022. [Online]. Available:
  \url{https://www.frontiersin.org/articles/10.3389/frcmn.2022.853416}
\BIBentrySTDinterwordspacing

\bibitem{f-FAMA}
K.-K. Wong and K.-F. Tong, ``Fluid antenna multiple access,'' \emph{IEEE Trans.
  Wireless Commun.}, vol.~21, no.~7, pp. 4801--4815, 2022.

\bibitem{COST2100}
J.~Flordelis, X.~Li, O.~Edfors, and F.~Tufvesson, ``Massive {MIMO} extensions
  to the {COST} 2100 channel model: Modeling and validation,'' \emph{IEEE IEEE
  Trans. Commun.}, vol.~19, no.~1, pp. 380--394, 2020.

\bibitem{138901}
\emph{Study on channel model for frequencies from 0.5 to 100 {GHz}}.\hskip 1em
  plus 0.5em minus 0.4em\relax 3GPP TR 38.901 version 18.0.0 (Release 18), Mar.
  2024.

\bibitem{NEURIPS2019_9015}
A.~Paszke, S.~Gross, F.~Massa, A.~Lerer, J.~Bradbury, G.~Chanan, T.~Killeen,
  Z.~Lin, N.~Gimelshein, L.~Antiga, A.~Desmaison, A.~Kopf, E.~Yang, Z.~DeVito,
  M.~Raison, A.~Tejani, S.~Chilamkurthy, B.~Steiner, L.~Fang, J.~Bai, and
  S.~Chintala, ``Pytorch: An imperative style, high-performance deep learning
  library,'' in \emph{Proc. Int. Conf. Neural Inf. Process. Syst.}, 2019, pp.
  8024--8035.

\bibitem{abadi2016tensorflow}
M.~Abadi, P.~Barham, J.~Chen, Z.~Chen, A.~Davis, J.~Dean, M.~Devin,
  S.~Ghemawat, G.~Irving, M.~Isard \emph{et~al.}, ``Tensorflow: A system for
  large-scale machine learning,'' in \emph{12th USENIX Symposium on Operating
  Systems Design and Implementation (OSDI 16)}, 2016, pp. 265--283.

\bibitem{cost2100@twc}
J.~Flordelis, X.~Li, O.~Edfors, and F.~Tufvesson, ``Massive {MIMO} extensions
  to the {COST} 2100 channel model: Modeling and validation,'' \emph{IEEE
  Trans. Wireless Commun.}, vol.~19, no.~1, pp. 380--394, 2020.

\end{thebibliography}

\end{document}